\documentclass[journal]{IEEEtran}
  \usepackage{amssymb}
  \usepackage{latexsym}
  \usepackage{amsfonts}
  \usepackage{amsthm}
  \usepackage{amsmath}
  \usepackage{graphics}
  \usepackage{setspace}
  \usepackage{graphicx}
  \usepackage{epstopdf}
  \usepackage[usenames,dvipsnames,svgnames,table]{xcolor}
  \usepackage{float}
  \usepackage{wrapfig}
  \usepackage{color}
  \usepackage{rotating}
  \usepackage{array} 
  \usepackage[hyphens]{url}
  \usepackage[table]{xcolor}
  \usepackage{multirow}  
  \usepackage{adjustbox}

  \usepackage{longtable}

  \usepackage{forloop,supertabular}
\newcounter{count}
\newcommand{\comment}[1]{}

\definecolor{SW}{RGB}{87,157,28}
\definecolor{SH}{RGB}{153,51,102}
\definecolor{SC}{RGB}{197,0,11}    
\definecolor{SE}{RGB}{0,132,209}
\definecolor{SN}{RGB}{255,149,14}

\newcommand{\catDxxWearable}{\adjustbox{valign=m}{\colorbox{SW}{}}}
\newcommand{\catExxSmartHome}{\adjustbox{valign=m}{\colorbox{SH}{}}}
\newcommand{\catAxxSmartCity}{\adjustbox{valign=m}{\colorbox{SC}{}}}
\newcommand{\catCxxEnvironment}{\adjustbox{valign=m}{\colorbox{SN}{}}}
\newcommand{\catBxxEnterprise}{\adjustbox{valign=m}{\colorbox{SE}{}}}

\begin{document}
%\title{Context-awareness in the Internet of Things Marketplace: A Survey\\ {\small Note: IEEE ACCESS Journal Official Title for Citations: 'A Survey on Internet of Things From Industrial Market Perspective'}}

\title{Context-aware Computing in the Internet of Things: A Survey on Internet of Things From Industrial Market Perspective\\ {\small Note: IEEE ACCESS Journal Official Title for Citations: 'A Survey on Internet of Things From Industrial Market Perspective'}}

%\title{A Survey on Internet of Things From Industrial Market Perspective: Focusing Context Awareness}

\author{Charith~Perera,~\IEEEmembership{~Member,~IEEE,}     
        Chi Harold Liu~\IEEEmembership{Member,~IEEE},
        Srimal~Jayawardena,~\IEEEmembership{~Member,~IEEE,}   
        Min Chen,~\IEEEmembership{~Senior Member, IEEE} 
%        \vspace{-0.8cm}% <-this % stops a space

\thanks{Charith~Perera, and Srimal~Jayawardena,  are with the Research School of Computer Science, The Australian National University, Canberra, ACT 0200, Australia. (e-mail: firstname.lastname@ieee.org)}% <-this % stops a space
\thanks{Chi Harold  Liu is with Beijing Institute of Technology, China. (e-mail: chiliu@bit.edu.cn)}
\thanks{Min Chen is with Huazhong University of Science and Technology, China. (e-mail: minchen@ieee.org)}

\thanks{Manuscript received xxx xx, xxxx; revised xxx xx, xxxx.}}


%\markboth{IEEE ACCESS,~Vol.~x, No.~x, xxxx~xxxx}%

\markboth{IEEE ACCESS}%
{Shell \MakeLowercase{\textit{et al.}}: Bare Demo of IEEEtran.cls for Journals}



\maketitle

\begin{abstract}
The Internet of Things (IoT) is a dynamic global information network consisting of Internet-connected objects, such as RFIDs, sensors, and actuators, as well as other instruments and smart appliances that are becoming an integral component of the Internet. Over the last few years, we have seen a plethora of IoT solutions making their way into the industry marketplace. Context-aware communication and computing has played a critical role throughout the last few years of ubiquitous computing and is expected to play a significant role in the IoT paradigm as well. In this article, we examine a variety of popular and innovative IoT solutions in terms of context-aware technology perspectives. More importantly, we evaluate these IoT solutions using a framework that we built around well-known context-aware computing theories.  This survey is intended to serve as a guideline and a conceptual framework for context-aware product development and research in the IoT paradigm. It also provides a systematic exploration of existing IoT products in the marketplace and highlights  a number of potentially significant research directions and trends.
\end{abstract}

\begin{IEEEkeywords}
Internet of things, industry solutions, context-awareness, product review, IoT marketplace
\end{IEEEkeywords}

%Then, we identify the trends as opportunities whereby context-aware technologies can be used to address open challenges
%\IEEEpeerreviewmaketitle

%.........................01. Introduction........................%

\section{Introduction}
\label{sec:Introduction}

Over the last few years the Internet of Things (IoT) \cite{P003} has gained significant attention from both industry and academia. Since the term was introduced in the late 1990s many solutions have been introduced to the IoT marketplace by different types of organization ranging from start-ups, academic institutions, government organizations and large enterprises \cite{ZMP007}. IoT's popularity is governed by both the value that it promises to create and market growth and predictions \cite{ZMP003}. The IoT allows '\textit{people and things to be connected Anytime, Anyplace, with Anything and Anyone, ideally using Any path/network and Any service}' \cite{P029}. Such technology will help to create '\textit{a better world for human beings}', where objects around us know what we like, what we want, and what we need and act accordingly without explicit instructions \cite{ZMP007}. 

Context-aware communication and computing is a key technology that enables intelligent interactions such as those which the IoT paradigm envisions. Let us briefly introduce some of the terms in this domain which will help to  better understand the remaining sections. Context can be defined \textit{as any information that can be used to characterize the situation of an entity. An entity is a person, place, piece of software, software service or object that is considered relevant to the interaction between a user and an application, including the user and application themselves} \cite{P104}. Context-awareness can be defined \textit{as the ability of a system to provide relevant information or services to users using context information where relevance depends on the user's task} \cite{P104}. Context-aware communication and computing has been researched extensively since the early 2000s and several surveys have been conducted in this field. The latest survey on context-aware computing focusing on the IoT was conducted by Perera et al. \cite{ZMP007}. Several other important surveys are analysed and listed in \cite{ZMP007}. However, all these surveys focus on academic research.

To the best of our knowledge, however, no survey has focused on industrial IoT solutions. All the above-mentioned surveys have reviewed the solutions proposed by the academic and research community and refer to scholarly publications produced by the respective researchers. In this paper, we review IoT solutions that have been proposed, designed, developed, and brought into the market by industrial organizations. These organizations range from start-ups and small and medium enterprises to large corporations. Because of their industrial and market-driven nature, most of the IoT solutions in the market are not published as academic work. Therefore, we collected information about the solutions from their respective websites, demo videos, technical specifications, and consumer reviews. Understanding how context-aware technologies are used in the IoT solutions in the industry's marketplace is vital for academics, researchers, and industrialists so they can identify trends, industry requirements, demands, and innovation opportunities. 

The rest of the article is organized as follows. In Section \ref{sec:Classification}, we briefly analyse IoT marketplace trends and growth. The evolution of context-aware technologies and applications are presented in Section \ref{sec:Evolution}. Then, we introduce the theoretical foundation and our evaluation framework used in this paper in Section \ref{sec:Theory}. Subsequently, in Section \ref{sec:Review}, we review a selected number of IoT solutions from context-aware perspective. Later, we present lessons learned and innovation opportunities based on the evaluation results in Section \ref{sec:Lessons_Learned}. Finally, we present the conclusion remarks.

\section{Internet of Things Marketplace}
\label{sec:Classification}

The vision of the IoT has been heavily  energised by statistics and predictions. In this section, we discuss some of the statistics and facts related to the IoT which allows us to understand how the IoT has grown over the years and how it is expected to grow in the future. Further, these statistics and facts highlight the future trends in the industry marketplace. 

\begin{figure}[!h]
 \centering
% \vspace{-0.43cm}
 \includegraphics[scale=0.48]{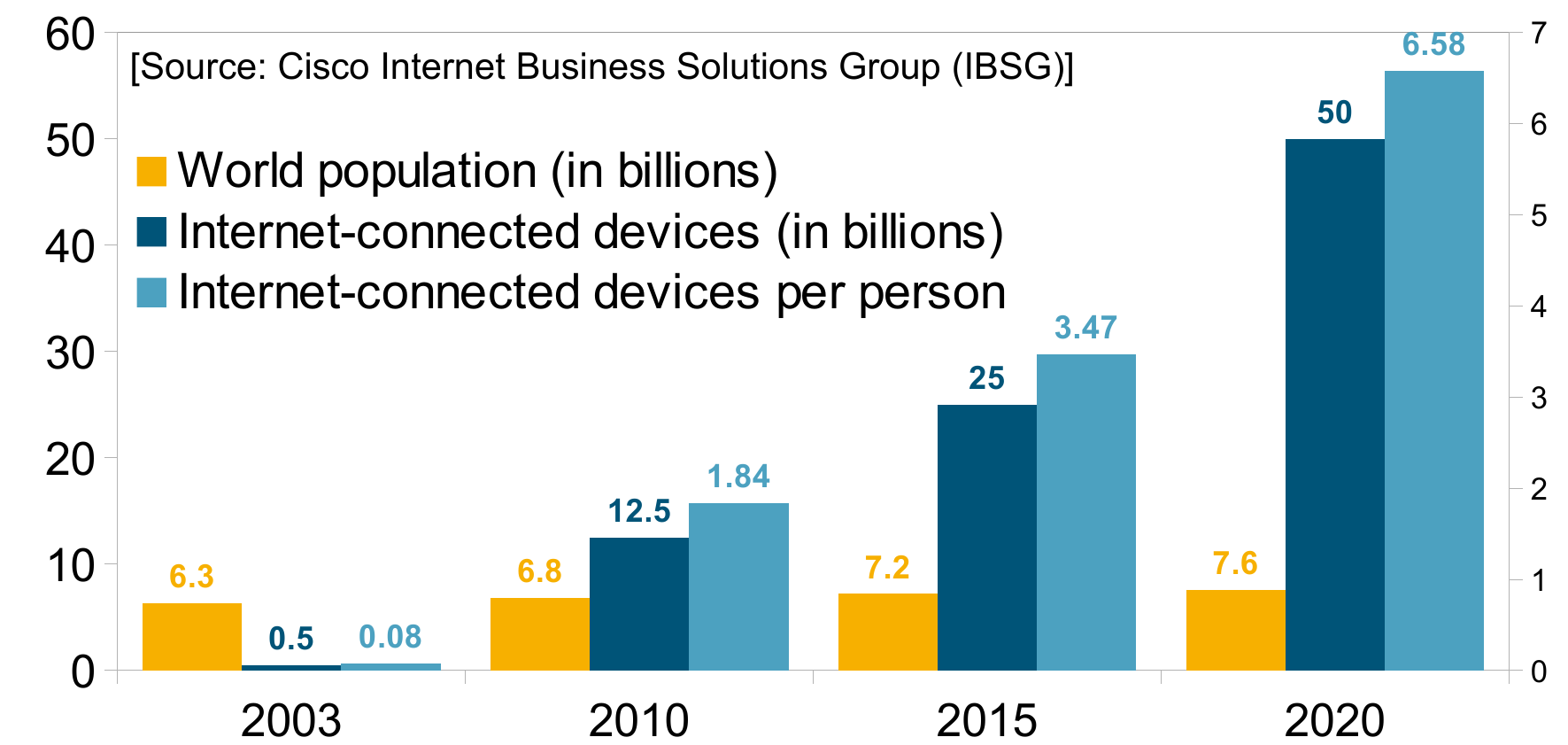}
%\vspace{-0.33cm}	
 \caption{Growth in Internet-Connected Devices / Objects by 2020.}
% \cite{Z1043}
 \label{Figure:Statistics3}	
%\vspace{-0.2cm}	
\end{figure}

It is estimated that there about 1.5 billion Internet-enabled PCs and over 1 billion Internet-enabled  mobile phones  today. These two categories  will  be joined by Internet-enabled smart objects \cite{P041, TII07}  in the future. By 2020, there will  be 50 to 100 billion devices connected to the Internet, ranging from smartphones, PCs, and ATMs (Automated Teller Machine) to manufacturing equipment in factories and products in shipping containers  \cite{ZMP008}. As depicted in Figure \ref{Figure:Statistics3}, the number of things connected to the Internet exceeded the number of people on Earth in 2008. According to CISCO, each individual on earth will have more than six devices connected to the Internet by 2020.

\begin{figure}[!h]
 \centering
% \vspace{-0.2cm}
 \includegraphics[scale=0.48]{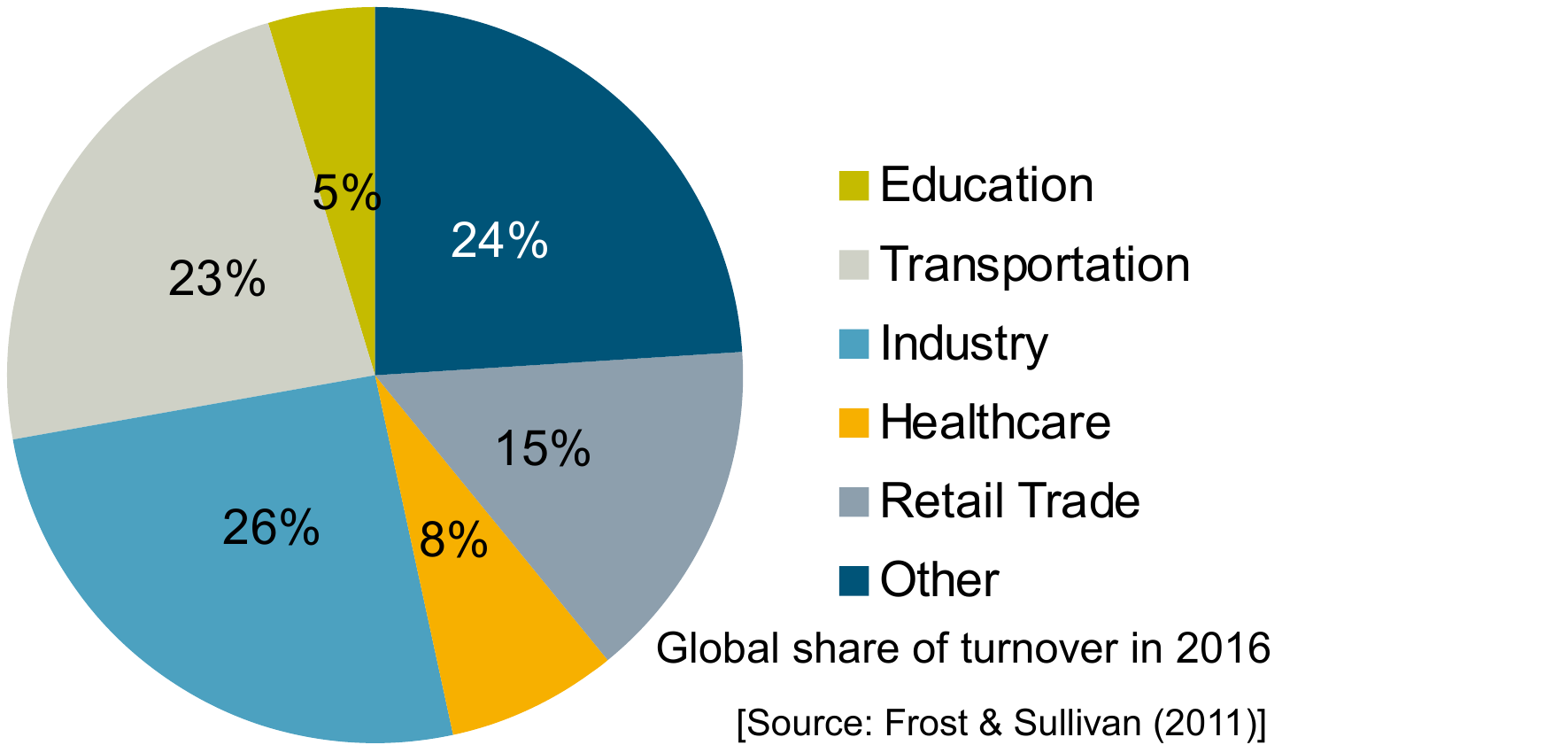}
%\vspace{-0.33cm}	
 \caption{RFID Sales by Major Market Segments.}
% \cite{Z1045}
 \label{Figure:Statistics2}	
%\vspace{-0.2cm}	
\end{figure}

According to BCC Research 2011 market report on sensors, the global market for sensors was around \$56.3 billion in 2010. In 2011, it was around \$62.8 billion. The global market for sensors is expected to increase to \$91.5 billion by 2016, at a compound annual growth rate of 7.8\%. One of the techniques for connecting everyday objects into networks is radio frequency identification — RFID technology \cite{Z1044}. In this technology, the data carried by the chip attached to an object is transmitted via wireless links. RFID has the capability to convert dump devices into comparatively smart objects. RFID systems can be used wherever automated labelling, identification, registration, storage, monitoring, or transport is required to increase efficiency and effectiveness. According to Frost \& Sullivan (2011), the global RFID market was valued at from \$3 billion to \$4 billion in 2009. The RFID market will grow by 20\% per year through 2016 and reach a volume of approximately from \$6.5 billion to almost \$9 billion. According to Figure \ref{Figure:Statistics2}, it is expected that five main sectors, education, transportation, industry, healthcare, and retails, will generate 76\% of the total RFID market demand by 2016.

\begin{figure}[!h]
 \centering
% \vspace{-0.43cm}
 \includegraphics[scale=0.48]{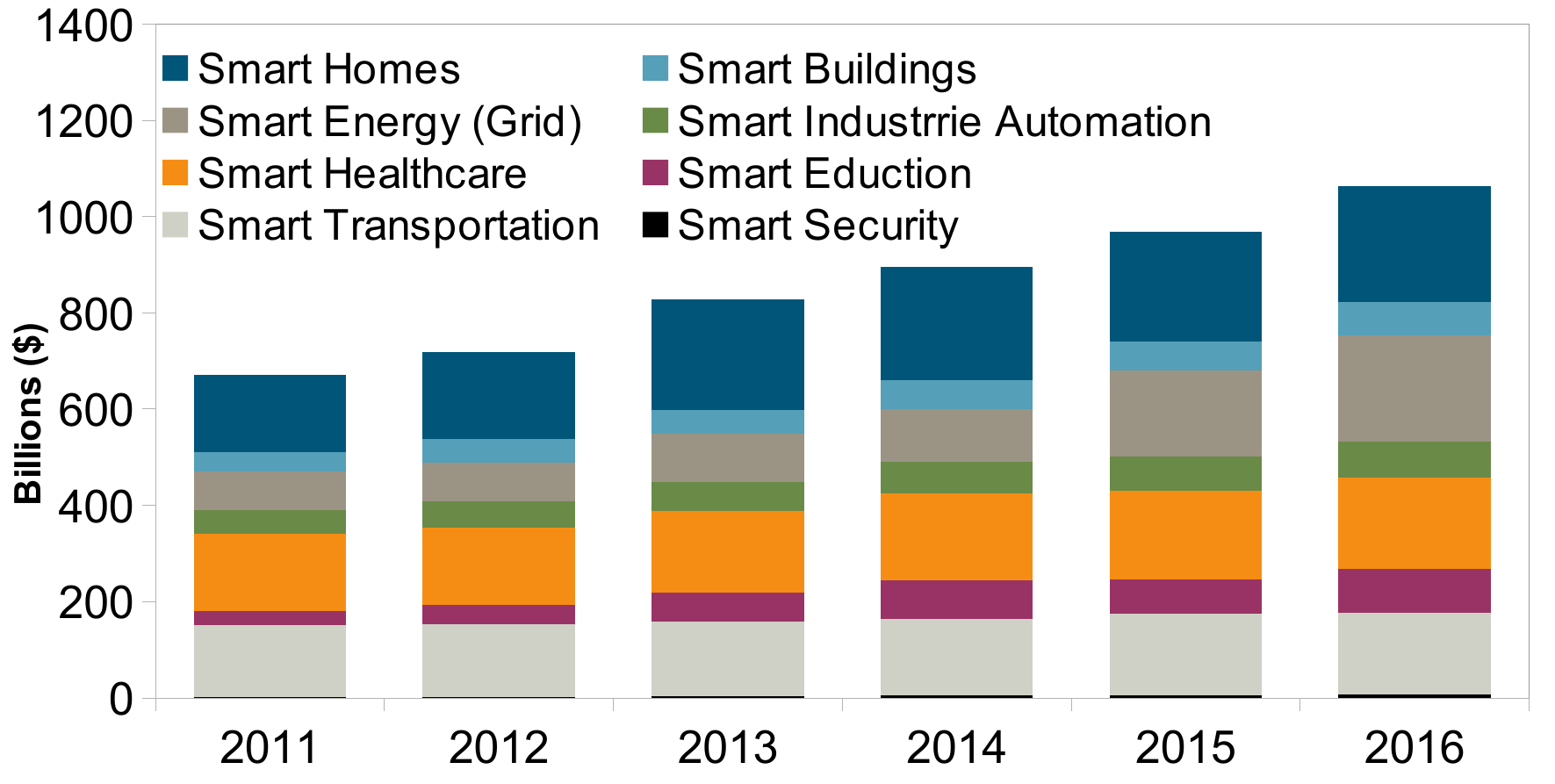}
%\vspace{-0.33cm}	
 \caption{Smart Product Sales by Market in 2016.}
 \label{Figure:Statistics1}	
%\vspace{-0.4cm}	
\end{figure}

 ``Smart city'' \cite{P532} is a concept aimed at providing a set of new generation services and infrastructure with the help of information and communication technologies (ICT). Smart cities are expected to be composed of many different smart domains. Smart transportation, smart security and smart energy management are some of the most important components for building smart cities \cite{TII04}. However, in term of market, smart homes, smart grid, smart healthcare, and smart transportation solutions are expected to generate the majority of sales.  According to MarketsandMarkets report on Smart Cities Market (2011 - 2016), the global smart city market is expected to cross \$1 trillion by 2016, growing at a CAGR of 14.2\% as illustrated in Figure \ref{Figure:Statistics1}.

The interconnection and communication between everyday objects, in the IoT paradigm, enables many applications in many domains. Asin and Gascon \cite{WaspMote} have listed 54 application domains under 12 categories: smart cities, smart environment, smart water, smart metering, security and emergencies, retail, logistics, industrial control, smart agriculture, smart animal farming, domestic and home automation, and eHealth. After analysing the industry marketplace and careful consideration, we classified the popular existing IoT solutions in the marketplace into five different categories: smart wearable, smart home, smart city, smart environment, smart enterprise. In this paper, we review  over 100 different IoT solutions in total. It is important to note that not all the solutions we examined are listed in the technology review in Table \ref{Tbl:Evaluation_of_Previous_Research_Efforts}. For the review, we selected a wide range of IoT products which demonstrate different context-aware functionalities.

%In Figure \ref{Figure:Evaluation_Methodology}, we illustrates our evaluation methodology where it shows how the total number of IoT solutions we surveyed are categorised under each category. . Because of space limitations, we review only the most popular IoT solutions in the market-place. Selection is also based on their application domain and functionality so we can cover a wider range of IoT solutions.

 \section{Evolution of Context-aware Technology}
 \label{sec:Evolution}

 \begin{figure*}[!t]
  \centering
 % \vspace{-0.43cm}
  \includegraphics[scale=0.65]{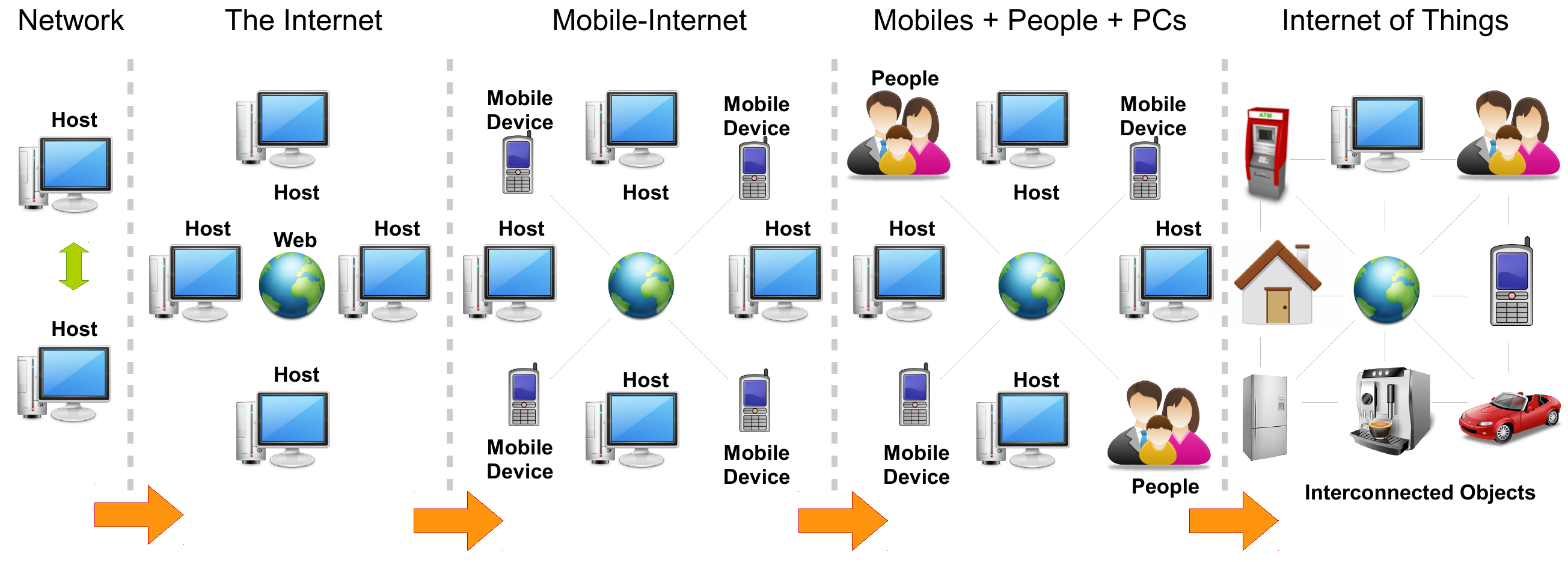}
 \vspace{-0.33cm}	
  \caption{Evolution of the Internet in five phases. The evolution of Internet begins with connecting two computers together and then moved towards creating  World Wide Web by connecting large number of computers together. The mobile-Internet emerged by connecting mobile devices to the Internet. Then, peoples'  identities joined the Internet via social networks. Finally, it is moving towards Internet of Things by connecting every day objects to the Internet}
  \label{Figure:Evolution_of_Internet}	
 \vspace{-0.4cm}	
 \end{figure*}

It is important to understand the evolution of the Internet before discussing the evolution of context-aware technologies. The Internet broadly evolved in five phases as illustrated in Figure \ref{Figure:Evolution_of_Internet}. The evolution of Internet begins with connecting two computers together and then moved towards creating the World Wide Web by connecting large number of computers together. Mobile-Internet emerged when mobile devices were connected to the Internet. People's identities were added to the Internet via social networks \cite{NP001}. Finally, the Internet of Things emerged, comprised of everyday objects added to the Internet. During the course of these phases, the application of context-aware communication and computing changed significantly \cite{ZMP007}.

In the early phase of computer networking when computers were connected to each other in point-to-point fashion, context-aware functionalities were not widely used. Providing help to users based on the context (of the application currently open) was one of the fundamental context-aware interactions provided in early computer applications and operating systems. Another popular use of context is context-aware menus that help users to perform tasks tailored to each situation in a given application. When the Internet came into being, location information started to become critical context information. Location information (retrieved through IP addresses) were used by services offered over the Internet in order to provide location-aware customization to users. Once the mobile devices (phones and tablets) became a popular and integral part of everyday life, context information collected from sensors built-in to the devices (e.g. accelerometer, gravity, gyroscope, GPS, linear accelerometer, and rotation vector, orientation, geomagnetic field, and proximity, and light, pressure, humidity and temperature) were used to provide context-aware functionality. For example, built-in sensors are used to determine user activities, environmental monitoring, health and well-being, location and so on \cite{P217}.

Over the last few years social networking \cite{IA01} has become popular and widely used. Context information gathered through social networking services \cite{IA08} (e.g. \textit{Facebook}, \textit{Myspace}, \textit{Twitter}, and \textit{Foursquare}) has been fused with the other context information retrieved through mobile devices to build novel context-aware applications such as activity predictions, recommendations, and personal assistance \cite{NP002}. For example, a mobile application may offer context-aware functionalities by fusing location information retrieved from mobile phones and recent `likes' retrieved from social media sites to recommend nearby restaurants that a user might like. In the next phase, `things' were connected to the Internet by creating the IoT paradigm. An example of context-aware functionality provided in the IoT paradigm would be an Internet-connected refrigerator telling users what is inside it, what needs to be purchased or what kind of recipes can be prepared for dinner. When the user leaves the office, the application autonomously does the shopping and guides the user to a particular shopping market so s/he can collect the goods it has purchased. In order to perform such tasks, the application must fuse location data, user preferences, activity prediction, user schedules, information retrieved through the refrigerator (i.e. shopping list) and many more. In the light of the above examples, it is evident that the complexity of collecting, processing and fusing information has increased over time. The amount of information collected to aid decision-making has also increased significantly.

 \section{Theoretical Foundation and Evaluation Framework}
 \label{sec:Theory}
 
 This section discusses context-aware theories and related historic developments over time. The evaluation framework which we used to review IoT products in the marketplace are built upon the theoretical foundation presented in this section. First, we lay the theoretical foundation and secondly we discuss the evaluation framework.

 \subsection{Context-aware Computing Theories}
 
 The term \textit{context} has been defined by many researchers. Dey et al.  \cite{P143} have evaluated and highlighted  the weaknesses of these definitions.  Dey claimed  that the definition provided by Schilit and Theimer  \cite{P173} was based on examples and cannot be used to identify new context. Further, Dey claimed that definitions provided by Brown  \cite{P175}, Franklin and Flachsbart \cite{P178}, Rodden et al. \cite{P181}, Hull et al. \cite{P179}, and Ward et al. \cite{P183} used synonyms  to refer to context,  such as `environment'   and `situation'.  Therefore,  these definitions  also cannot be used to identify new context. Abowd and Mynatt  \cite{P115} have identified the five W's (Who, What, Where, When, Why) as the minimum  information that is necessary to understand context. Schilit et al.  \cite{P116} and Pascoe \cite{P180} have also defined the term context. 
 
 We accept the  definition of context provided by Abowd et al. \cite{P104} to be used  in this research  work, because  their definition can be used to identify context from data in general. We presented the definition of \textit{context} in Section \ref{sec:Introduction}.
 
% \textit{``Context is any information that can be used to characterise the situation of an entity. An entity is a person, place, or object that is considered relevant to the interaction between a user and an application, including the user and applications themselves \cite{P104}.''} 
 
 \begin{figure}[!b]
  \centering
 % \vspace{-0.43cm}
  \includegraphics[scale=0.74]{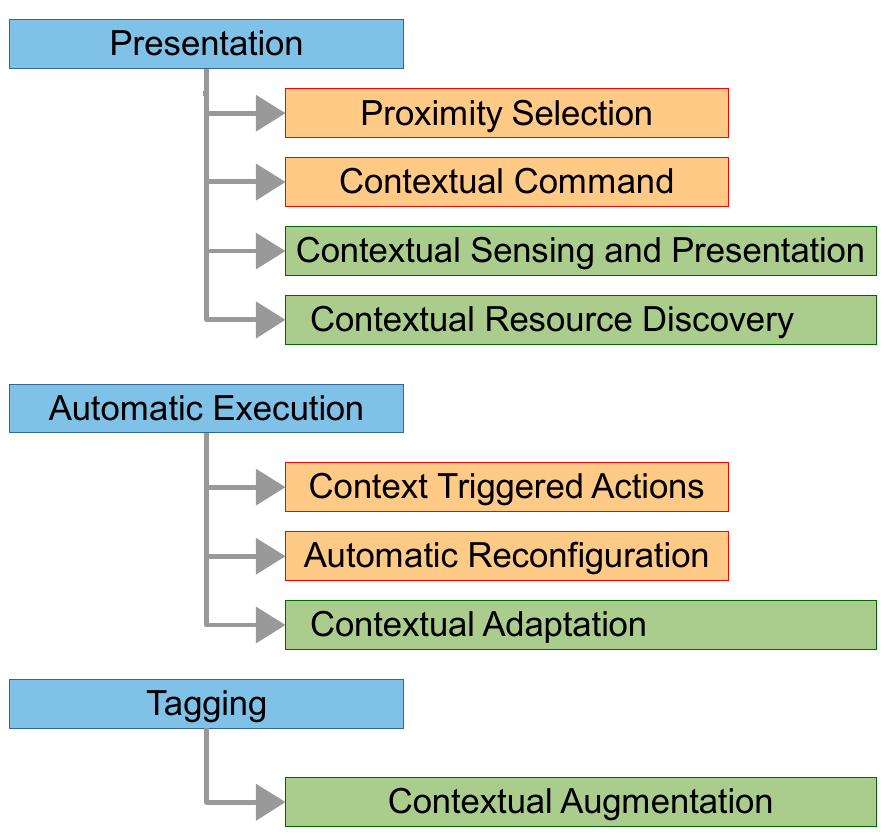}
 %\vspace{-0.33cm}	
  \caption{Context-aware features identified by different researchers: Abowd et al.  \cite{P104} (Blue), Schilit et al.  \cite{P116} (Yellow), Pascoe \cite{P180} (Green). Ccontext-awareness as been defined using these features (can also be called characteristics of a given system) }
  \label{Figure:Theoratical_Summary}	
 %\vspace{-0.4cm}	
 \end{figure}

 The term \textit{context awareness}, also called \textit{sentient}, was first introduced by Schilit and Theimer  \cite{P173}  in 1994. Later, it was defined  by Ryan et al.  \cite{P182}. In both cases, the focus was on computer applications and systems. As stated by Abowd et al.  \cite{P104}, those definitions  are too specific and cannot be used to identify whether a given system is a context-aware  system or not. We presented the definition provided by Abowd et al.  \cite{P104} in Section \ref{sec:Introduction}.  After analysing and comparing the two previous efforts conducted  by Schilit et al.  \cite{P116} and Pascoe \cite{P180}, Abowd et al. \cite{P104} identified three features that a context-aware  application  can support:  presentation,  execution, and tagging. Even though, the IoT vision was  not known at the time these  features  are identified,  they are highly applicable to the IoT paradigm  as well. We elaborate these features from an IoT perspective.
 
% Therefore, Abowd et al.  has defined the term context-awareness  as follows:
 
% \textit{``A system is context-aware if it uses context to provide relevant information and/or services to the user, where relevancy depends on the user's task. \cite{P104}''}

 \begin{itemize}
\item \textbf{Presentation:} Context  can be used to decide what information  and services need to be presented to the user. Let us consider a smart \cite{P007} environment  scenario. When a user enters a supermarket   and takes their smart phone out, what they want to see is their shopping list. Context-aware mobile  applications  need to connect to kitchen appliances  such as a smart refrigerator \cite{P352} in the home to retrieve  the shopping list and present it  to the user. This provides the idea of presenting information  based on context such as location,  time, etc. By definition, IoT promises to provide any service anytime, anyplace, with anything and anyone, ideally using any path/network.

 \item \textbf{Execution:} Automatic execution  of services  is also a critical feature in the IoT paradigm. Let us consider  a smart home  \cite{P007} environment. When a user  starts driving home from their office, the IoT application employed  in  the house should switch  on the air condition system and switch on the coffee machine  to be ready to use  by the time the user  steps into their house. These actions need to be taken automatically  based on the context. Machine-to-machine communication is a significant  part of the IoT.

 \item\textbf{Tagging:} In the IoT paradigm,  there will be a  large number of sensors attached to everyday objects. These objects will produce  large volumes  of sensor  data that has  to be collected, analysed,  fused and interpreted \cite{P109}. Sensor data produced by a  single sensor  will  not provide the necessary information that can be used to fully understand the situation \cite{IA06}.  Therefore,  sensor data collected  through multiple sensors  needs  to be fused together \cite{IA05}.  In order to accomplish the sensor data fusion task, context needs to be collected.  Context needs to be tagged together with the sensor data to be processed and understood later. Context annotation  plays a  significant role in context-aware computing research. The  \textit{tagging} operation also identified as \textit{annotation}. 

 \end{itemize}

   \begin{figure*}[!t]
    \centering
   % \vspace{-0.43cm}
    \includegraphics[scale=0.68]{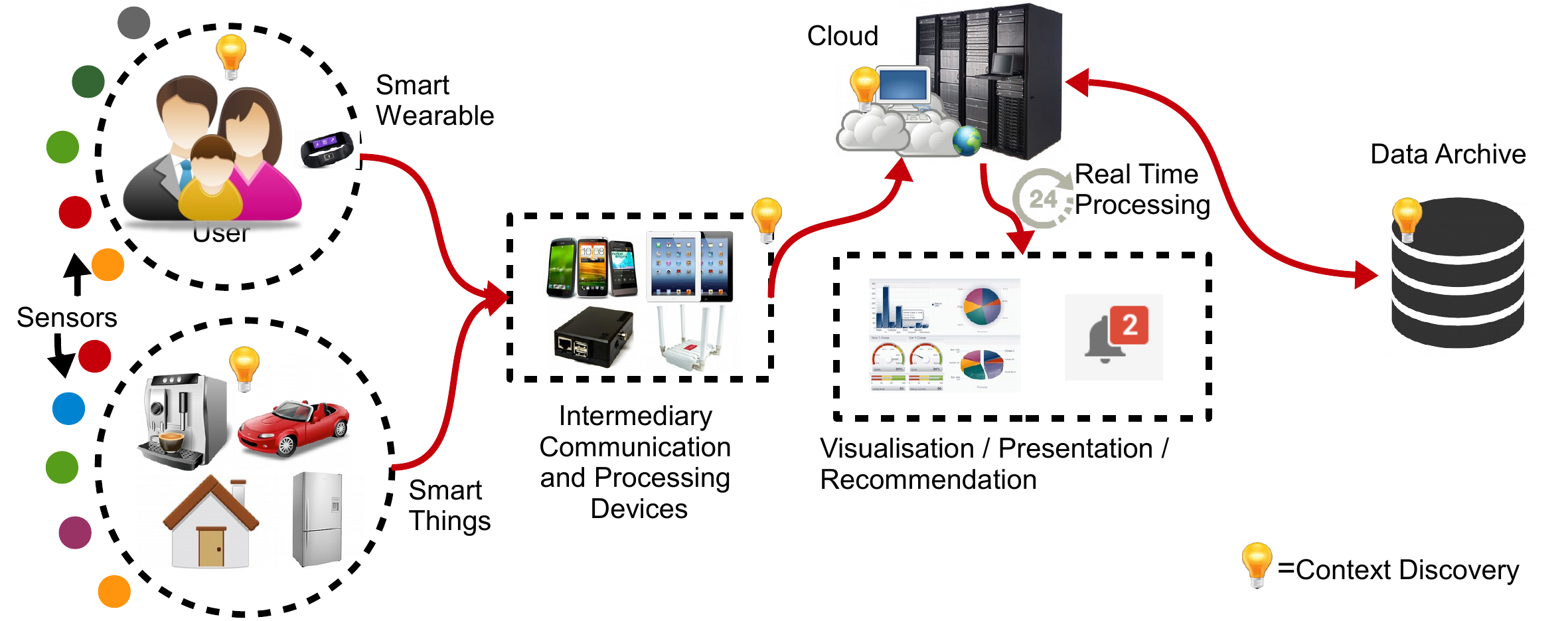}
   \vspace{-0.33cm}	
    \caption{Data Flow in IoT Solutions in High-level. Context can be discovers in different stages / phases in the data flow. A typical IoT solution may use some part of the data flow architecture depending on the their intended functionalities.}
    \label{Figure:Context_aware_Big_Picture}	
   \vspace{-0.4cm}	
   \end{figure*}

% After analyzing and comparing the two previous efforts conducted by Schilit et al. \cite{P116} and Pascoe \cite{P180}, Abowd et al. \cite{P104} identified three features that a context-aware application can support: presentation, execution, and tagging. Even though, the IoT vision was not known at the time these features are identified, they are highly applicable to the IoT paradigm as well. The features are 1) presentation of information and services to a user; 2) automatic execution of a service; and 3) tagging of context to information for later retrieval. Further, these features are discussed in IoT perspective in \cite{ZMP007}.

  In Figure \ref{Figure:Theoratical_Summary}, we summarise three different context-aware features presented by  researchers. It is clear that all these classification methods have similarities. We have considered all these feature sets when developing our evaluation framework.

  \subsection{Evaluation Framework}

This section presents the evaluation framework we used to review the IoT products in context-aware perspective. We developed this evaluation framework based on the widely recognized and cited research done by Abowd et al. \cite{P104}.  In this evaluation, we apply one and half decade old context aware theories into IoT era. Our evaluation is mainly based on three context-aware features in high-level: 1)\textit{ context-aware selection and presentation}, 2) \textit{context-aware execution}, and 3) \textit{context-aware-tagging}.  However, we have also enriched the evaluation framework by identifying sub-features under above mentioned three features. Our evaluation framework consists of nine (9) features.

% First, we briefly outline the original theoretical foundation presented by Abowd et al. Secondly, we present hot these theories can be used to evaluate industrial solutions in IoT the market place from context-aware perspective. 

The  Figure \ref{Figure:Context_aware_Big_Picture} visualizes how data is being collected transferred, processed, context  discovered and annotated in typical IoT solutions. It is important to note that not all solutions may use the exact same data flow. Each solution may use part of the architecture in their solution. We will refer to this common data flow architecture during this paper to demonstrate how each solution may design their data flows. Our objective is to identify major strategies that are used by IoT products to offer context-aware functionalities. From here onwards, we explain the taxonomy, the evaluation framework, used to evaluate the IoT products. The results of the evaluation are presented in Table \ref{Tbl:Evaluation_of_Previous_Research_Efforts}. Summary of the evaluation framework is presented in Table \ref{Tbl:Summarized_taxonmy}.
 
First we introduce the name of the IoT solution in the column (1) in Table \ref{Tbl:Evaluation_of_Previous_Research_Efforts}. We also provide the web page link of the each product / solution. It is important to note that, these products does not have any related academic publication. Therefore, we believe that web page links are the most reliable reference to a given IoT solution. Such links allow readers to follow further reading by using the product name along with web link.

In column (2), we classify each product into  five categories. Each category is denoted by a different colour: \iftrue red {\color{SC}\rule{0.2cm}{0.2cm}}   (smart city), yellow 
{\color{SN}\rule{0.2cm}{0.2cm}}  (smart environment), blue {\color{SE}\rule{0.2cm}{0.2cm}}  (smart enterprise), green  {\color{SW}\rule{0.2cm}{0.2cm}} (smart wearable), and purple {\color{SH}\rule{0.2cm}{0.2cm}} (smart home). \else red \catC  (smart city), yellow \catV   (smart environment), blue \catN  (smart enterprise), green \catW (smart wearable), and purple \catH (smart home). \fi Some solutions may belong to multiple categories. We  divide the rest of the columns into three section : \textit{Context-aware Tagging}, \textit{Context Selection and Presentation}, and \textit{Context execution}.

\subsubsection{Context-aware Tagging Section}

 \begin{figure}[!b]
  \centering
  \vspace{-0.43cm}
  \includegraphics[scale=0.45]{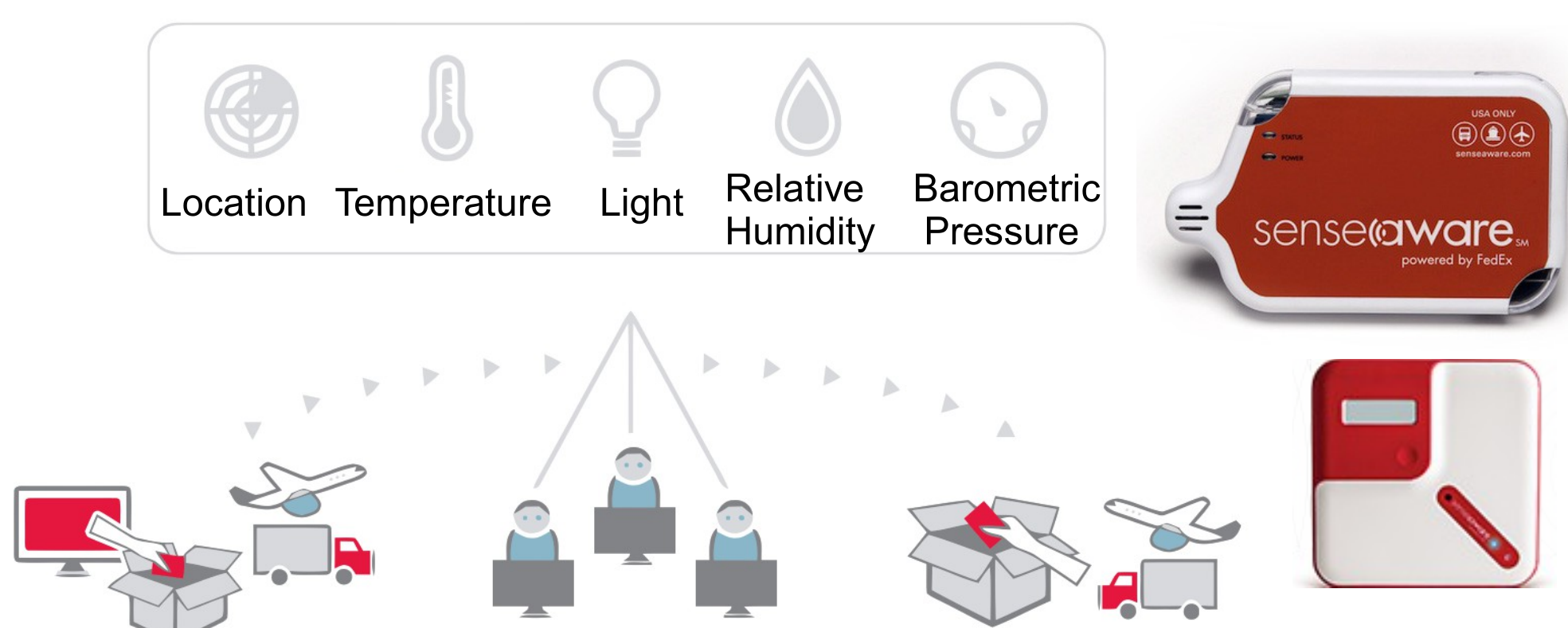}
 %\vspace{-0.33cm}	
  \caption{\textit{SenseAware} (senseaware.com) uses small smart devices that comprises five different built-in sensors with limited computational and communication capabilities. It reports the status of the packages in real time to the cloud. These smart devices comes in different sizes and form factors, as illustrated here, in order to support different types of packaging methods (Two types of smart devices are shown in the figure)}
  \label{Figure:Primary_Context}	
 %\vspace{-0.4cm}	
 \end{figure}

Context-aware tagging, which is also called context augmentation and annotation represent the idea of sensing the environment and collecting primary context information. We also believe that secondary context generation is also a part of context-aware tagging feature. Primary context is any information retrieved without using existing  context and without performing any kind of sensor  data fusion operations \cite{ZMP007}.  For example, \textit{SenseAware} (senseaware.com) is a solution developed to support real-time shipment tracking. As illustrated in Figure \ref{Figure:Primary_Context}, \textit{SenseAware} collects and processes context information such as location, temperature, light, relative humidity and biometric pressure  in order to enhance the visibility and transparency of the supply chain.  \textit{SenseAware} uses both hardware and software components in their sensor-based logistic solution. such  data collection allows different parties engage in supply chain to monitor the movement of goods in real-time and accurately know the quality of the transported goods and plan their processes effectively and efficiently. We list commonly acquired primary context information in column (3) in Table \ref{Tbl:Evaluation_of_Previous_Research_Efforts}.

    \begin{figure}[!b]
     \centering
     \vspace{-0.43cm}
     \includegraphics[scale=0.55]{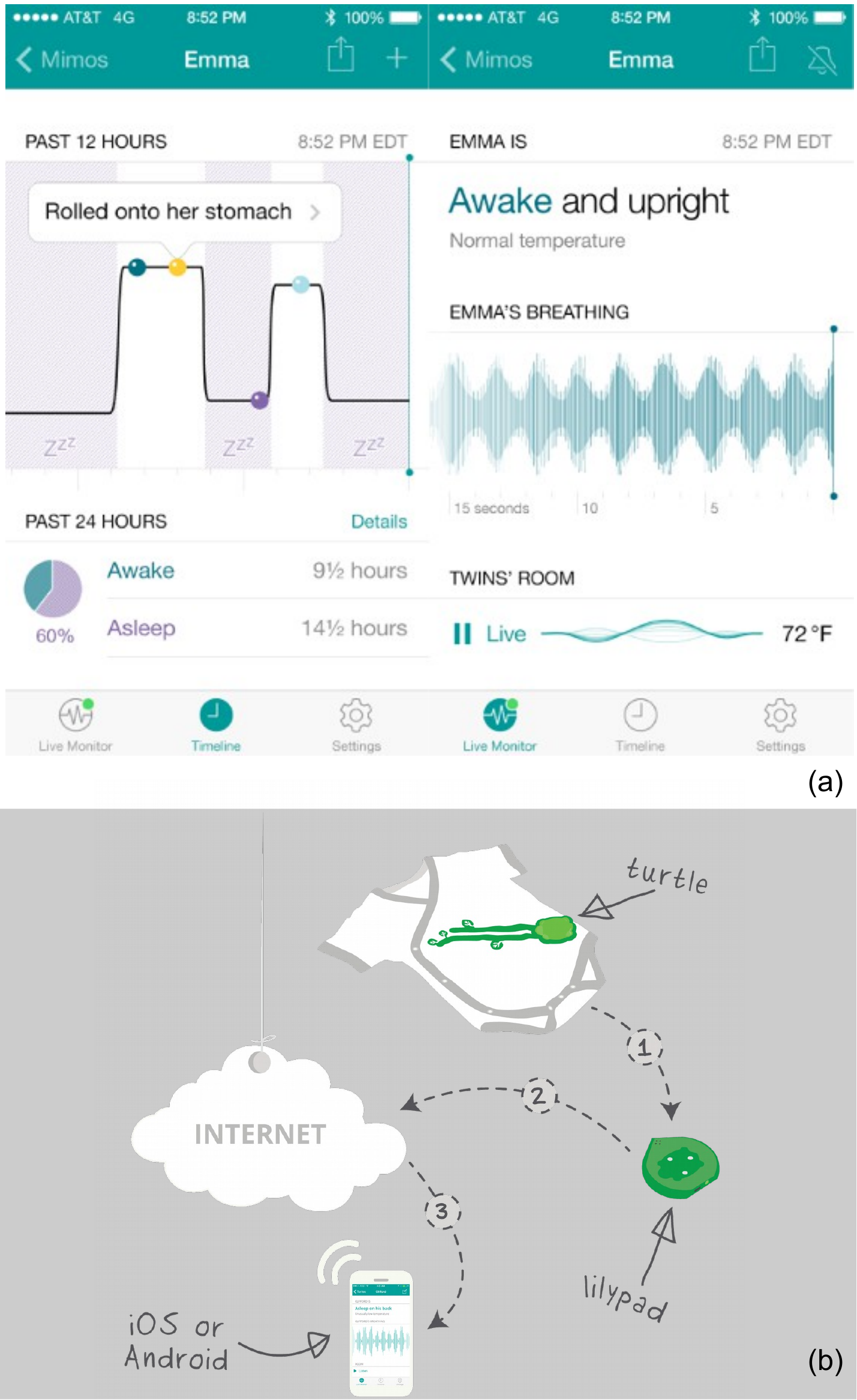}
    %\vspace{-0.33cm}	
     \caption{(a) User interface provided to the users, in this case parents by \textit{Mimo Smart Baby Monitor} (mimobaby.com). All the raw information collected are presented to the users, using graphs, figures and icons, after generating secondary context information. (b) Illustrates how primary context has been collected and transferred through the infrastructure to discover secondary context information.}
     \label{Figure:Secondary_Context}	
    %\vspace{-0.4cm}	
    \end{figure}
 
Secondary context is any information that can be computed using primary context. The secondary context can be computed by using sensor data fusion operations or data retrieval operations such as web service calls (e.g. identify the distance between two sensors by applying sensor data fusion operations on two raw GPS sensor values). Further, retrieved context such as phone numbers, addresses, email addresses, birthdays, list of friends from a contact  information provider based on a personal identity as the primary  context can also be identified  as secondary context.  For example, \textit{Mimo} (mimobaby.com) has built a smart nursery system, where parents learn new insights about their baby through connected products like the \textit{Mimo Smart Baby Monitor}. In this product, \textit{turtle} is the device that collects all primary context information. Then the data is transferred to an intermediary devices called \textit{lilypad}. Such responsibility offloading strategy allows to reduce the \textit{turtle}'s weight at minimum level and to increase the battery life. the communication and processing capabilities are offloaded to the \textit{lilypad} device which can be easiy recharged when necessary. We can see \textit{Mimo Smart Baby Monitor} usees some parts of the data flow architecture we presented in Figure \ref{Tbl:Evaluation_of_Previous_Research_Efforts}. User interface provided by Mimo and the data flow within the solution is presented in Figure \ref{Figure:Secondary_Context}. Cloud services \cite{IA02} performs the additional processing and summarised data is pushed to the mobile devices for context presentation. In the user interface, parents are presented mostly the secondary context information such as baby movement or baby's sleeping status. Accelerometer sensors are used to discover such secondary context information using pattern recognition techniques. We list secondary context information generated by IoT solutions in column (4) in Table \ref{Tbl:Evaluation_of_Previous_Research_Efforts}.
 
  \begin{figure}[!b]
   \centering
  % \vspace{-0.43cm}
   \includegraphics[scale=0.50]{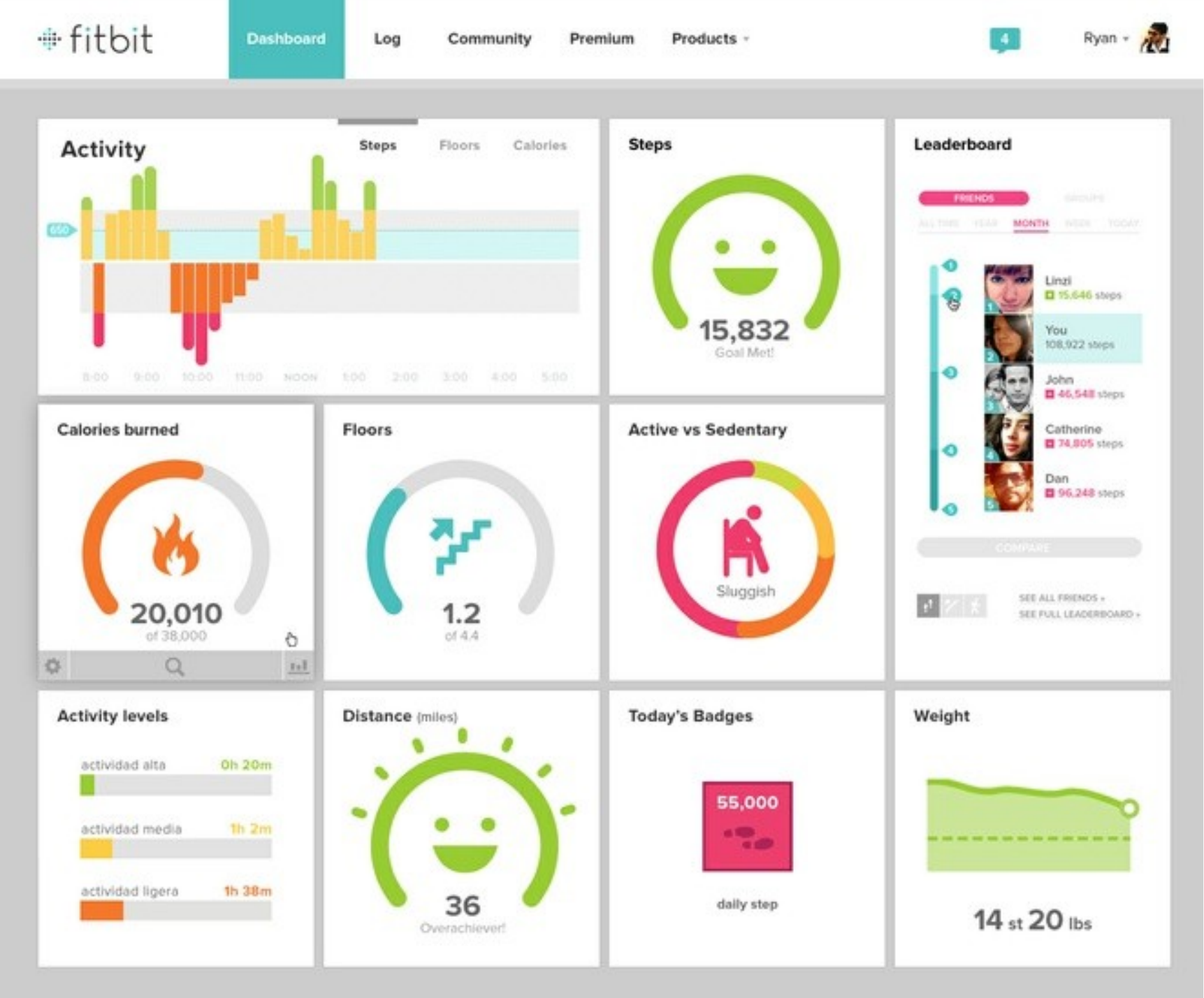}
  %\vspace{-0.33cm}	
   \caption{The \textit{Fitbit} web based dashboard displays recent activity level and lots of other statistics using graphics, charts, and icons.}
   \label{Figure:Context_Presentation}	
  %\vspace{-0.4cm}	
  \end{figure}
 
\subsubsection{Context Selection and Presentation Section}

There are number of commonly used strategies, by most of the IoT solutions in the marketplace, to present context to the users.
 
Most of the IoT products use some kind of visualization techniques to present context information the users. We call this \textit{visual presentation}. For example, \textit{Fitbit}  (fitbit.com) is a device that can be worn on multiple body parts in order to tracks steps taken, stairs climbed, calories burned, and hours slept, distance travelled, quality of sleep. This device collects data and present it to the users through mobile devices and web interfaces. Figure \ref{Figure:Context_Presentation} illustrates the context presentation of \textit{Fitbit}. Variety of different charts, graphs, icons and other types of graphical elements are heavily used to summarise and present analysed meaningful actionable data to the users. such visualization strategies are commonly encouraged in human computer interaction domain specially due to the fact that \textit{'a picture is worth a thousand words'}. We denote the presence of virtual presentation related to each IoT product using (\checkmark) in column (5) in Table \ref{Tbl:Evaluation_of_Previous_Research_Efforts}.

\begin{figure}[!h]
    \centering
   % \vspace{-0.43cm}
    \includegraphics[scale=0.36]{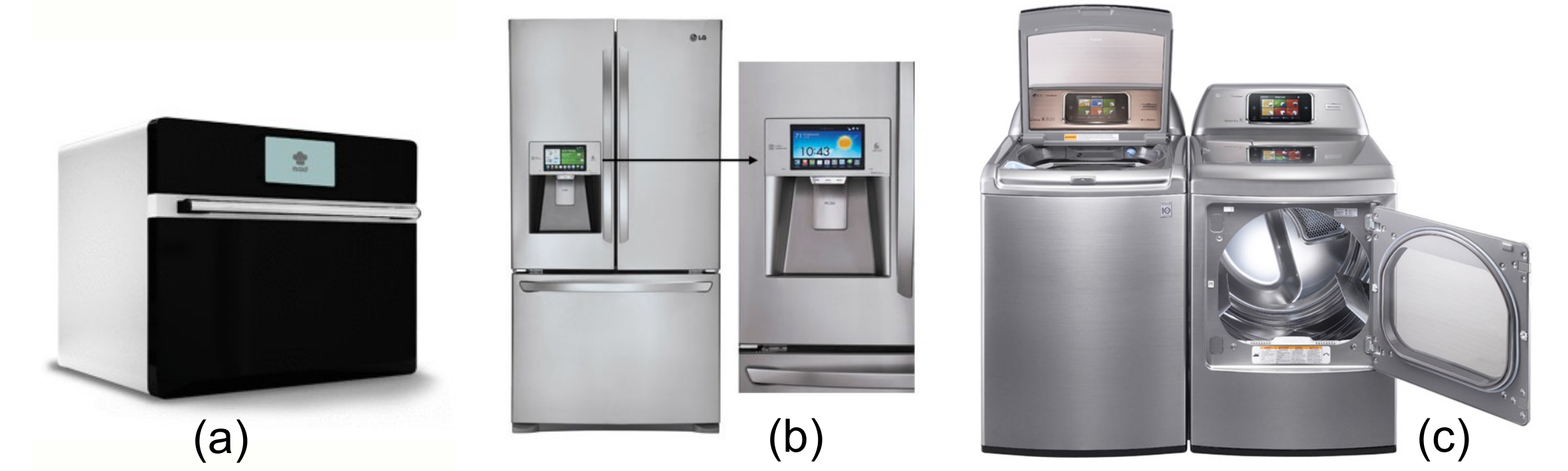}
   %\vspace{-0.33cm}	
    \caption{(a) Smart Oven (maidoven.com), (b) Smart Fridge (lg.com/us/discover/smartthinq/refrigerator), (c) Smart Washing machine (lg.com/us/discover/smartthinq/laundry). Some of the commonly used objects in households are not enriched with presentation capabilities such as touch screens. In such circumstances  context selection and presentation responsibilities can be offloaded to commonly used devices such as smart-phones and tablets.}
    \label{Figure:Object_based}	
   %\vspace{-0.4cm}	
   \end{figure}

IoT solutions in the market place also employ different commonly used devices to present the context to the users. Typically, an IoT solution offers context presentation and selection via some kind of software application. Some of the commonly used presentation channels are web-based (W), mobile-based (M), desktop-based  (D), and objects-based (O). First, three mediums describes themselves. Object-based means that context selection and presentation is done through  a custom IoT device itself. Sample IoT solutions that use object-base presentation strategy are presented in Figure \ref{Figure:Object_based}. We identify the presence of different  presentation channels related to each IoT product  in column (6) in Table \ref{Tbl:Evaluation_of_Previous_Research_Efforts}.

In addition to the context presentation  channels, IoT solutions use number of user interaction mechanisms such as voice (V), gesture (G), touch (T). Over last few years, we have seen more and more voice activated IoT solutions are coming to the marketplace. For example, latest technological development such as natural language processing and semantic technologies have enabled the wide use of  voice activated IoT solutions. \textit{Amazon Echo} (amazon.com/oc/echo), \textit{Ubi} (theubi.com) are  two voice activated personnel assistant solutions.  Typically, they are capable of answering user queries related to weather, maps, traffic and so on (i.e. commonly asked questions). They are designed to learn from user interactions and customize their services and predictive models based on the user behaviour and preferences. These solutions have gone beyond what typical smart phone assistants such as \textit{Google One}, \textit{Microsoft Cortana}, \textit{Apple Siri} has to offer. For example, \textit{Ubi} has the cabability to interact with other smart objects in the smart house environment.

More importantly products such as \textit{Ivee} (helloivee.com), a voice controlled hub for smart homes, facilitates interoperability over the other IoT products in the markets. This means that consumers can use \textit{Ivee} to control other IoT products \textit{Iris} (irissmarthome.com), \textit{Nest} (nest.com), \textit{Philips Hue} (meethue.com), \textit{SmartThings} (smartthings.com), and Belkin WeMo (belkin.com). We discuss interoperability matters in details in Section \ref{sec:Lessons_Learned}. In addition to centralizes home hubs based IoT systems, more and more standalone IoT products also support voice-activated interaction such as executing commands. For example, \textit{VOCCA} (voccalight.com) is a  plug \& play voice activated light bulb adapter requires no WiFi, no set-up, no installation. 

Gesture has also been used to enable the interactions between IoT products and users. For example \textit{Myo} (thalmic.com/en/myo/) is a wearable armband that can be used to issues gesture base commands. \textit{Myo} reads gestures and motion and let hte users to seamlessly control smart phones, presentations, and so on. \textit{Nod} (hellonod.com) is the a advanced gesture control ring. It allows users to engage objects with user movements. Nod can be considered as a universal controller, allowing effortless communication with all of the smart devices in users connected life, including phones, tablets, \textit{Google Glass}, watches, home appliances, TVs, computers and more. We identify the presence of different user interaction mechanisms related to each IoT product  in column (7) in Table \ref{Tbl:Evaluation_of_Previous_Research_Efforts}.

IoT solutions process data in different locations in their data communication flow as shown in Figure \ref{Figure:Context_aware_Big_Picture}. Sometimes data is processed within the sensors or the local processing devices. In other circumstances, data is sent to the cloud for processing. Deepening the applications and functionalities each IoT solution tries to provide, data may be processed in real-time (RT) or later (A). Specially, event detection based IoT systems need to act in real-time which requires real-time processing. For example, IoT solutions such as \textit{Mimo} smart baby monitor performs data processing in real-time as their mission is to increase the health and safety of the toddlers. It is also important to note that not every solution requires data archival. For example, health and fitness related IoT products can be benefited from archiving historic data. Such archives data will allow to produce graphs and charts over time and provide   more insights and recommendations t the consumers. More data also facilitates more accurate prediction. However. storing more data cost more and not every solution requires such storage. \textit{ShutterEaze} (shuttereaze.com)  makes it easy for anyone to add remote control functionality and automate their existing interior plantation shutters. For example, IoT product like this will  not necessarily  be benefited by  archiving historic data. Still it can learn user behaviour over time (based on how users use the product), and automate the task without storing data. We identify the usage of real-time and archival techniques   in column (8) in Table \ref{Tbl:Evaluation_of_Previous_Research_Efforts}.

IoT solutions mainly use three different reaction mechanisms. Most commonly used mechanism is notification (N). This means that when a certain condition is met, IoT solution will release a notification to the users explaining the context. For example, \textit{Mimo} (mimobaby.com), the baby monitoring product we mentioned earlier, notifies the parents when the baby shows any abnormal movements or breathing patterns. Parent will receive the notification through their smart phone. Some IoT solutions may react by performing actuations (A). For example, \textit{Blossom} (myblossom.com) ia a smart watering products that can be self-programmed based on real-time weather data and gives the user control over the phone, lowering the water bill up to 30\%. In this kind of scenario, the product may autonomously perform the actuations (i.e. open and close sprinklers) based on the context information. Another reaction mechanism used by IoT solutions is providing recommendations (R). For example, \textit{MAID} (maidoven.com)  has a personalization engine that continuously learns about the users. MAID learns what users cook regularly, tracks users activity using data from smart phones and smart watches. Then, it will  provide recommendations for a healthy balanced diet. \textit{MAID} also recommends users to workout or to go for a run based on the calories they consume each day. We identify the usage of reaction mechanisms related to each IoT product  in column (9) in Table \ref{Tbl:Evaluation_of_Previous_Research_Efforts}.

 %%%%%%%%%%%%%%%%%%%%%%%%%%%%%%%%%%%
\begin{table*}[t]
\centering
\footnotesize

\caption{Summary of the Evaluation Framework Used in Table~\ref{Tbl:Evaluation_of_Previous_Research_Efforts}} 
\vspace{-0.3cm}
\begin{tabular}{ c l m{13cm} }
\hline
 & Taxonomy / Feature & Description \\ \hline \hline
%1 & Project Name &  \\ 
%2 & Citations &  \\ 
%3 & Year &  \\ 
%4 & Project Focus &  \\ 
%
1 & Product and Web link & The name of the IoT product or the solution  sorted by `Category' and then by `Project Name' within each category in ascending order.\\ 
2 & Category & Category that the solution belongs to. 
Each category is denoted by a different colour: 
\iftrue
red {\color{SC}\rule{0.2cm}{0.2cm}}   (smart city), yellow 
{\color{SN}\rule{0.2cm}{0.2cm}}  (smart environment), blue {\color{SE}\rule{0.2cm}{0.2cm}}  (smart enterprise), green  {\color{SW}\rule{0.2cm}{0.2cm}} (smart wearable), and purple {\color{SH}\rule{0.2cm}{0.2cm}} (smart home).
\else
red \catC  (smart city), yellow \catV   (smart environment), blue \catN  (smart enterprise), green \catW (smart wearable), and purple \catH (smart home).
\fi
Some solutions belongs to multiple categories.\\ 
3 & Primary Context & Major context data  captured by IoT solutions. 
%We use following abbreviations to denote different types of context data: Temperature (T) \textcolor{red}{to be added...} 
\\ 
%% Check all converted to USD?
4 & Secondary Context & Major secondary context  generated by the IoT solution. 
%We use following abbreviations to denote different types of secondary context information: Activity (A) \textcolor{red}{to be added...}  
\\
5 & Visual Presentation & We denote  the presence of visual context presentation using a (\checkmark). \\
6 & Presentation Channels &   We identify a number of commonly used presentation channels as follows: Web-based (W), Mobile-based (M), Desktop-based (D), Object-based (O). Please note that web based channels can be accessed through both mobile and desktop devices. However, we consider web-based as a separate category while native mobile apps considered as mobile based and native desktop apps consider as desktop-based.\\
7 & User Interaction Mechanism &  We identify Touch (T), Gesture (G), and Voice (V) as three commonly used user interaction mechanism. Interactions done through a PC or a smart phone is denoted by (M). Touch (T) refers to the \textit{`user touching a physical product'}. It does not refer to the user interaction using touch enabled devices such as smart phones.\\
8 & Real-Time or  Archival & Some IoT solutions processes data in real-time (RT) and other process archival data (A). \\
9 & Reaction  Mechanism &  IoT products use different reaction mechanisms. Some of them release notifications (N). Some solutions provides recommendation (R) to the users on how to react to a certain situation. Some IoT products perform physical actuations (A).\\
10 & Learning Ability & Some solutions are capable of learning  by analysing user behaviours and other inputs over time. such machine leaning ability is denoted by (ML). Other solutions require specific instruction from users typically using IF-ELSE-THEN mechanism. Such user defined approach is denoted using (UD). \\
11 & Notification Execution &   In IoT products, notifications are released based in different conditions as follows: Temporal (T), Spatial (S), Event (E). Notification could be in any form such as SMS, email, sound, vibration and so on.\\
%12 & User Involvement &  Some IoT Products can work auto nomously (A). Other solutions requre actions to be tabken by users (M). \\ 

  \\ \hline
\multicolumn{3}{c}{Note: Cases where sufficient information were not available are denoted by (-). Further, ($\times$) denote the unavailability of a certain feature.}
\end{tabular}
\label{Tbl:Summarized_taxonmy}
%\vspace{-16pt}
\end{table*}
%%%%%%%%%%%%%%%%%%%%%%%%%%%%%%%%%%%

\begin{figure}[!h]
       \centering
      % \vspace{-0.43cm}
       \includegraphics[scale=0.65]{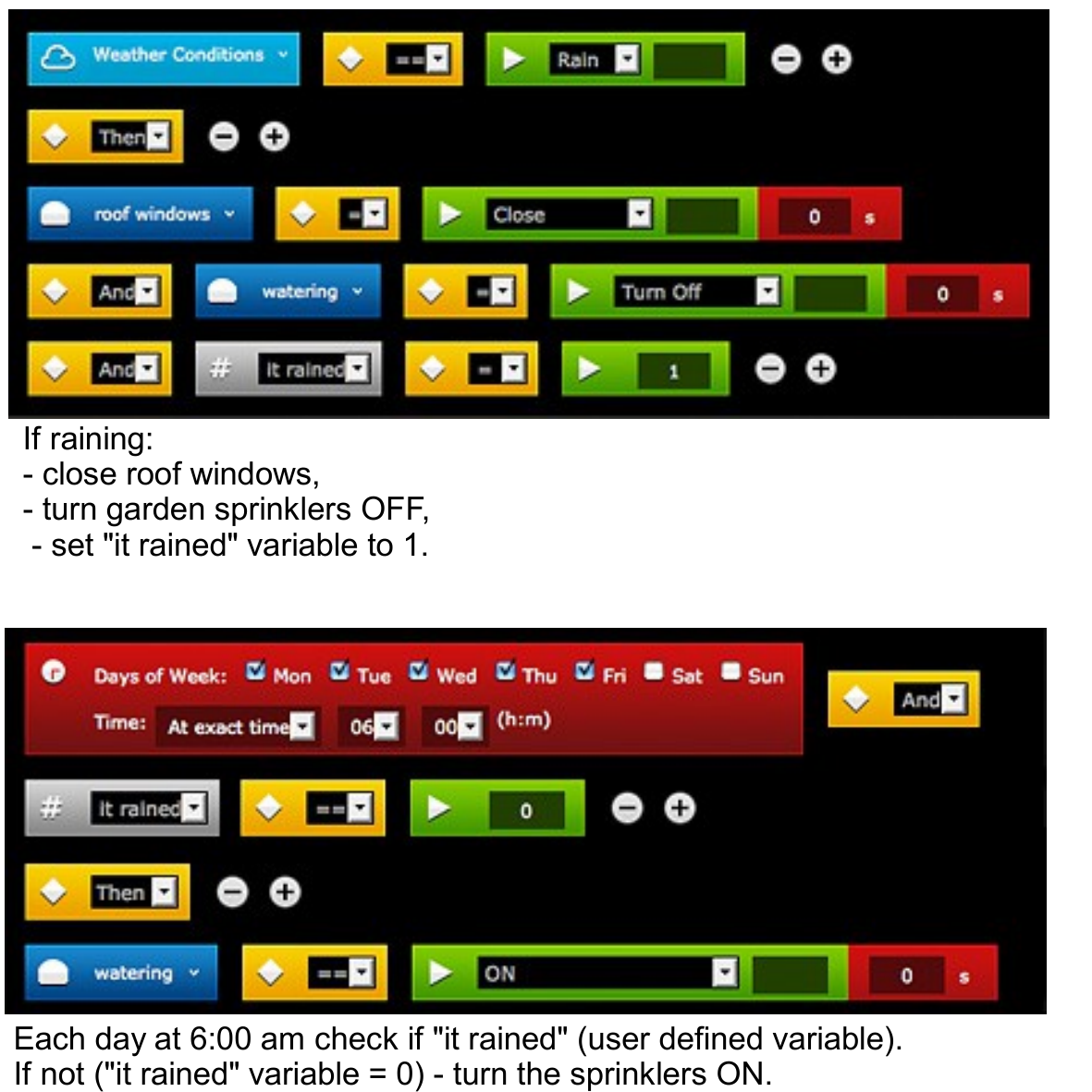}
      %\vspace{-0.33cm}	
       \caption{Two scenarios defined using \textit{Fibaro} (fibaro.com) platforms. The screen-shots  show how different types of context triggered can be defined by combining sensors, actuators and predefined parameters.}
       \label{Figure:Manual_Execution}	
      %\vspace{-0.4cm}	
      \end{figure}

Another important factor we identified during the product review is the learn-ability. Some products are capable of recording user provided inputs and other autonomously gathered information to predict future behaviours. In computer science, such behaviour is identified as machine learning (ML). For example, \textit{Nest} (nest.com) thermostat is capable of learning users' schedules and the temperatures users prefer. It keeps users comfortable and saves energy when they are away. In contrast, products such as \textit{Fibaro} (fibaro.com) requires users to explicitly  defines (UD) event thresholds and triggers as shown in Figure \ref{Figure:Manual_Execution}. We review the learn-ability of each  IoT product  in column (10) in Table \ref{Tbl:Evaluation_of_Previous_Research_Efforts}.

\begin{figure}[!b]
        \centering
       % \vspace{-0.43cm}
        \includegraphics[scale=0.34]{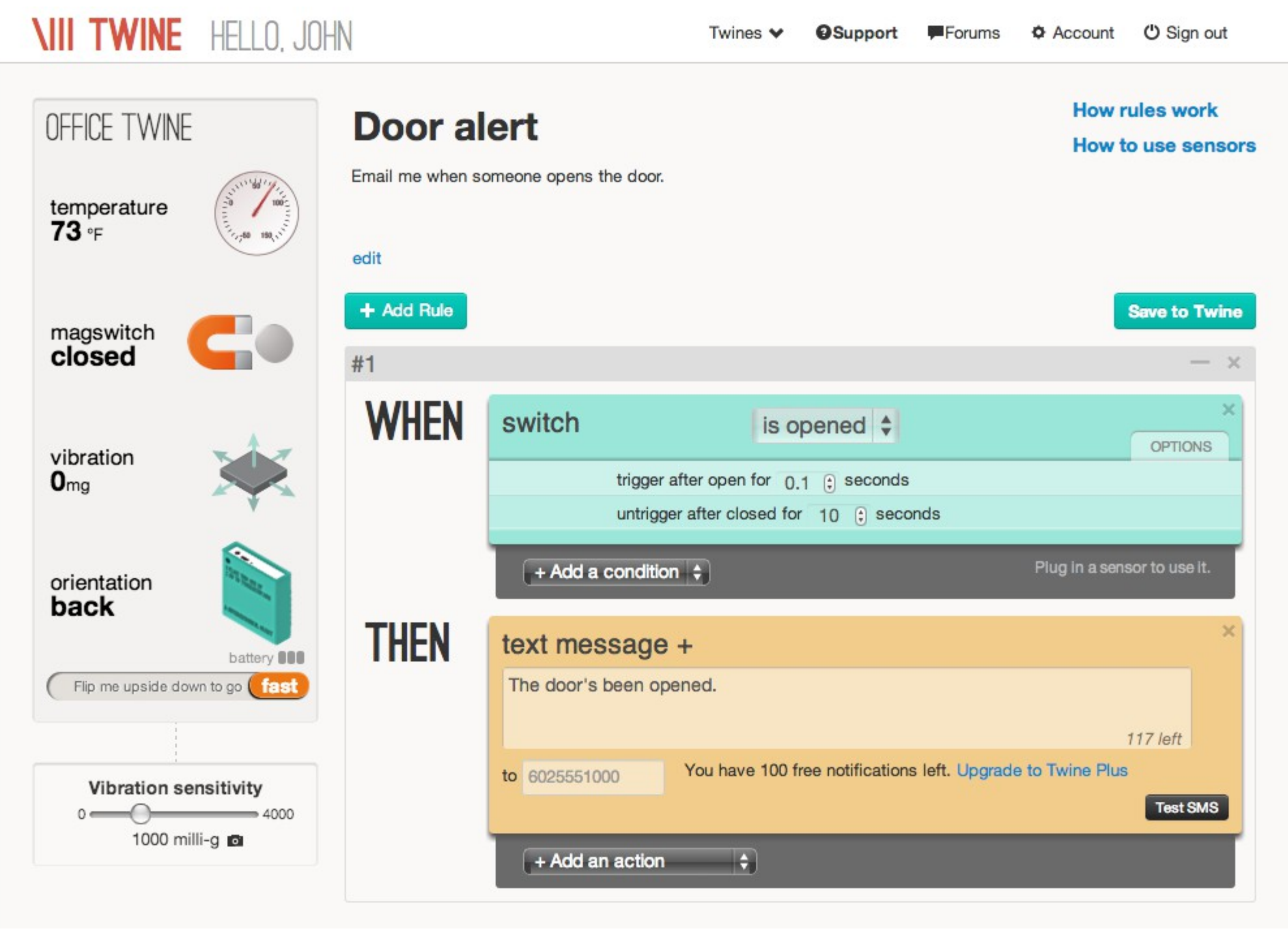}
       %\vspace{-0.33cm}	
        \caption{\textit{Twine} (supermechanical.com) provides a user interface to define scenarios  by combining sensors and actuators in a WHEN-THEN fashion which is also similar to the IF-THEN mechanism. \textit{Twine}  will trigger the actuation accordingly when conditions are met.}
        \label{Figure:Event_Trigger}	
       %\vspace{-0.4cm}	
       \end{figure}

There are number of different ways that an IoT product would trigger a certain reaction. It is important to note that a single IoT solution may combine multiple triggers together in order to facilitate complex requirements. Some rigger may be spacial (S), temporal (T),  or event based (E). Event based triggers are the most commonly used mechanism. For example, the IoT products such as \textit{SmartThings} (smartthings.com), \textit{Ninja Blocks} (ninjablocks.com), \textit{Fibaro} (fibaro.com), \textit{Twine} (supermechanical.com) allow users to define contextual triggers using sensors, actuators and parameters. Figure \ref{Figure:Manual_Execution} and Figure \ref{Figure:Event_Trigger} shows how two different products define events.

Low powered bluetooth beacons are commonly used in IoT products, specially in commercial and retail sector for both localization and location-based advertising \cite{IA002}.   For example, \textit{XY} (xyfindit.com) and \textit{Estimote} (estimote.com) are two similar products in the IoT marketplace that provide small beacons that  can be attached to any location or object. The beacons will broadcast tiny radio signals which  smart phones can receive and interpret, unlocking micro-location and contextual awareness. Therefore, IoT products may trigger a reaction when either users entering into or going out from a certain area. There are some other products such as \textit{FiLIP}  (myfilip.com) which users location-aware triggers to make sure children are staying within safe area. \textit{FiLIP}  uses a unique blend of GPS, GSM, and WiFi to allow parents to locate their child using the most accurate location information, both indoors and outdoors. Parents can create a virtual radius around a location, such as home, school or a friend's house. Further, parents can set up to five such safe zones using the \textit{FiLIP} app. A notification will be sent to the parent's smart phone when \textit{FiLIP}  detects that the child has entered or left a safe zone.

In temporal  mechanism, trigger is release based on a time schedule. Temporal triggers may refer to time as time of the day (e.g. exactly: 10.30 am or approximately: morning), day of the week (e.g. Monday or weekend), week of the month (e.g. second week), month of the year (e.g January), season (e.g. winter). Figure \ref{Figure:Manual_Execution} show how \textit{Fibaro} system allows to define a trigger by incorporating temporal triggers. IoT products such as \textit{Nest} thermostat also use temporal triggers to efficiently learn and manager energy consumption.

  \section{Review of IoT Solutions}
  \label{sec:Review}

In this section we evaluated variety of different IoT solutions in the marketplace based on the evaluation framework presented in the earlier section. Table \ref{Tbl:Summarized_taxonmy}, summarises the evaluation framework used and Table \ref{Tbl:Evaluation_of_Previous_Research_Efforts} presents the IoT product review results.

%============================= Table 1===================================
%========================================================================

\begin{table*}[t!]
\caption{Evaluation of Surveyed Research Prototypes, Systems, and Approaches}
\footnotesize
\renewcommand{\arraystretch}{0.8}
\begin{tabular}{
 p{2.5cm} 
 c 
 m{2.8cm}  
 m{2.8cm}
 c
 c
 c
 c
 c
 c
 c 
 } 
%  \hline 
%  \multicolumn{3}{|c|}{OneTwoThree} \\
%  \\
 \hline
 
  \begin{minipage}[b]{2.0cm}[Project Type] Project Name (Web Link) \end{minipage}      &          %--(1)
 \begin{sideways}Category \end{sideways}   &   %--(2)
 \begin{minipage}[b]{2.0cm}Primary Context \end{minipage}  &                                         
 \begin{minipage}[b]{2.2cm}Secondary Context \end{minipage}  &
 \begin{sideways}\begin{minipage}[b]{1.4cm}Visual \\Presentation\end{minipage} \end{sideways} & %--(7)
 \begin{sideways}\begin{minipage}[b]{1.4cm}Presentation Channel\end{minipage} \end{sideways} & %--(7)
 \begin{sideways}\begin{minipage}[b]{1.4cm}User \\Interaction\end{minipage} \end{sideways} & %--(7)
 \begin{sideways}\begin{minipage}[b]{1.4cm}Real-Time \\ Archival \end{minipage} \end{sideways} &  %--(8)
 \begin{sideways}\begin{minipage}[b]{1.4cm}Notification \\ Mechanism \end{minipage} \end{sideways} &   \begin{sideways}\begin{minipage}[b]{1.4cm}Learning Ability \end{minipage} \end{sideways} &  %--(8)
\begin{sideways}\begin{minipage}[b]{1.4cm}Notification Execution \end{minipage} \end{sideways}   %--(8)

 \\  
 \hline \hline 
% \vspace{0.3cm}  
(1)      & (2)              & (3)      & (4)      &(5)            & (6)       & (7)      & (8)    & (9)                            &  (10)   &  (11)                 \\
%   BASIS    & \cite{basis}     & 2012     &  W       & $\checkmark$  & 199$^{1}$   & H,S      & W, B   & A$^{M}$, I$^{M}$, B$^{S}$      &  C      &  U      & P               \\

%\renewcommand{\arraystretch}{1.05}

%-------------------------------- From spreadsheet - Begin
[Waste Management] Enevo (enevo.com) & \catAxxSmartCity & Waste fill-level & Efficient routes to pick-up waste,  schedules &  \checkmark & W & M & RT, A & N, R & ML, UD & E  \\  

[Indoor Localization] Estimote (estimote.com) & \catAxxSmartCity & Bluetooth signal strength, Beacon ID & Location, Distance &  \checkmark & M & M & RT & N, R & UD & T, S, E  \\  

[Parking Slot Management] ParkSight (streetline.com) & \catAxxSmartCity & Sound level, Road surface temperature & Route for free parking slot &  \checkmark & M, W & M & RT, A & N, R & ML, UD & T, S, E  \\  

[Street Lighting] Tvilight (tvilight.com) & \catAxxSmartCity & Light, Presence, Local information such as weather changes, special events, emergency situations & Energy   consumption, Energy usage patterns, Lamp failure detection &  \checkmark & W & M & RT, A & N, A & ML, UD & T, S, E  \\  

[Crowed Movement Analysis] SceneTap (scenetap.com) & \catAxxSmartCity & GPS, Video & Crowd profiling at a given location &  \checkmark & M, W, D & M & RT & N, A & ML & T, S  \\  

[Foot Traffic Monitoring] Scanalyticsinc (scanalyticsinc.com) & \catAxxSmartCity & Floor level & Heat maps to understand customer movements &  \checkmark & W & T, M & RT, A & N & ML, UD & S, E  \\

[Crowed  Analysis] Livehoods (livehoods.org) & \catAxxSmartCity &  \textit{Foursquare} check-ins cloud service & Social dynamics, structure, and character of cities on  large scale &  \checkmark & W & M & RT, A & - & ML & E  \\  

[Crowed  Analysis] Placemeter (placemeter.com) & \catAxxSmartCity & Location, Video & Crowed movement &  \checkmark & M & M & RT & - & ML & E  \\  

[Fire Safety] Fire Extinguishers (engaugeinc.net) & \catBxxEnterprise & Pressure gauge, Motion  & Fire extinguisher usage patterns, Storage quality  &  \checkmark & W & M & RT, A & N & UD & S, E  \\  

[Foot Traffic Monitoring]  Motionloft (motionloft.com) & \catBxxEnterprise & Video, Motion & Location, Movement direction, Predict pedestrian and vehicle traffic &  \checkmark & W & M & RT, A & N & ML, UD & E  \\  

[Indoor Localization] Museum Analytics (artprocessors.net) & \catBxxEnterprise & Bluetooth signal strength, Beacon ID & Location, Distance &  \checkmark & W & M & RT & N & UD & S, E  \\  

[Supply Chain Management] SenseAware (senseaware.com) & \catBxxEnterprise & GPS, Temperature, Humidity, Light, Pressure & Shipment quality &  \checkmark & W & M & RT, A & N & UD & S, E  \\

[Manufacturing Process Management] Sight Machine (sightmachine.com) & \catBxxEnterprise & Video, Mechanical movements of Robots & Quality and efficiency of  manufacturing operations &  \checkmark & W & M & RT, A & N & ML, UD & E  \\  

[Concrete Structure Health Monitoring] Smart Structures (smart-structures-inc.us) & \catBxxEnterprise & Accelerometers, Strain gages, Temperature & Real-time load capacity, construction quality &  \checkmark & D & M & RT, A & N & UD & E  \\  

[Smart Pallet] (igps.net) & \catBxxEnterprise & RFID, Barcode & Identify item using Global Returnable Asset Identifier (GRAI) & - & W & M & RT & N & - & S, E  \\  

[Order Picking Glass] SmartPick (smartpick.be) & \catBxxEnterprise & Video, Barcode & Identify products, Identify the tasks to perform related to each object &  \checkmark & O & T, G & RT & N, R & - & S, E  \\  

[Environmental Monitoring] AirCasting (aircasting.org) & \catCxxEnvironment & Sound levels, Temperature, Humidity, CO, NO$_{2}$    & Air quality maps &  \checkmark & O, M, W & T, M & RT, A & N, R & UD, ML & E  \\  

\hline
\label{Tbl:Evaluation_of_Previous_Research_Efforts}

\end{tabular}
\end{table*}

%============================= Table 2===================================
%========================================================================

\begin{table*}[t!]

\footnotesize
\renewcommand{\arraystretch}{0.8}
\begin{tabular}{
 p{2.5cm} 
 c 
 m{2.8cm}  
 m{2.8cm}
 c
 c
 c
 c
 c
 c
 c 
 } 
%  \hline 
%  \multicolumn{3}{|c|}{OneTwoThree} \\
%  \\
 \hline
 
  \begin{minipage}[b]{2.0cm}[Project Type] Project Name (Web Link) \end{minipage}      &          %--(1)
 \begin{sideways}Category \end{sideways}   &   %--(2)
 \begin{minipage}[b]{2.0cm}Primary Context \end{minipage}  &                                         
 \begin{minipage}[b]{2.2cm}Secondary Context \end{minipage}  &
 \begin{sideways}\begin{minipage}[b]{1.4cm}Visual \\Presentation\end{minipage} \end{sideways} & %--(7)
 \begin{sideways}\begin{minipage}[b]{1.4cm}Presentation Channel\end{minipage} \end{sideways} & %--(7)
 \begin{sideways}\begin{minipage}[b]{1.4cm}User \\Interaction\end{minipage} \end{sideways} & %--(7)
 \begin{sideways}\begin{minipage}[b]{1.4cm}Real-Time \\ Archival \end{minipage} \end{sideways} &  %--(8)
 \begin{sideways}\begin{minipage}[b]{1.4cm}Notification \\ Mechanism \end{minipage} \end{sideways} &   \begin{sideways}\begin{minipage}[b]{1.4cm}Learning Ability \end{minipage} \end{sideways} &  %--(8)
\begin{sideways}\begin{minipage}[b]{1.4cm}Notification Execution \end{minipage} \end{sideways}   %--(8)
% \begin{sideways}\begin{minipage}[b]{1.6cm}User \\Involvement \end{minipage} \end{sideways}    %--(12) 
 \\  
 \hline \hline 
% \vspace{0.3cm}  
(1)      & (2)              & (3)      & (4)      &(5)            & (6)       & (7)      & (8)    & (9)                            &  (10)   &  (11)                 \\

[Air Quality Monitor] Air Quality Egg (airqualityegg.com) & \catCxxEnvironment & NO$_{2}$, CO, O$_{3}$, Volatile Organic Compounds, Radiation, Dust particulars & Air quality maps &  \checkmark & W & M & RT, A & N & UD & E  \\  

[Public Sensor Infrastructure] Array of Things (arrayofthings.github.io) & \catCxxEnvironment & Temperature, Humidity, Light, CO ,NO$_{2}$,  vibration, Volatile organic compounds, O$_{3}$, CO$_{2}$, SO,  Dust particulars, Sound, infra-red images,  Precipitation and wind measurements  & Climate trends, Air quality & - & M & M & RT, A & - & - & E  \\  

[Smart Farming] Bumblebee  project (niksargent.com/bumblebee) & \catCxxEnvironment & Video, Audio, Temperature, Sunlight, Weather & Model bees' life styles and behaviour &  \checkmark & D & - & A & N & - & -  \\  

[Smart River Management] Floating Sensor Network (float.berkeley.edu) & \catCxxEnvironment &  GPS, Temperature, Salinity & Maps of water movement, Hydrodynamic modelling. &  \checkmark & W & M & RT, A & N & ML & E  \\  

[Floot Detection] Oxford Flood Network (oxfloodnet.co.uk) & \catCxxEnvironment & Temperature, Ultrasonic, Wet sensor & Flood detection and prediction &  \checkmark & W & M & RT, A & N & ML, UD & E  \\  

[Weather Monitor] PressureNet (pressurenet.cumulonimbus.ca) & \catCxxEnvironment & Barometer, GPS & Weather Forecast &  \checkmark & M & M & RT, A & N & ML, UD & E  \\  

[Waste Management] Smart Belly (bigbelly.com) & \catCxxEnvironment & Waste fill-level & Efficient routes to pick-up waste &  \checkmark & M, W & M & RT, A & N, R & ML, UD & S, E  \\  

[Environment Monitor] Tzoa (mytzoa.com) & \catCxxEnvironment & Air Quality, UV, Temperature, Humidity, Light & Air Quality in streets, Indoor air quality maps  &  \checkmark & O, M & M & RT, A & N, R & UD & E  \\  

[Weather Monitor] (uniform.net) & \catCxxEnvironment & - & Retrieve weather information from Web a service & $ \times $ & O & T & RT, A & N, A & ML & E  \\  

[Sleep Monitor] Beddit (beddit.com) & \catDxxWearable & Force sensor, Heart rate sensor & Heart rate, Respiration, Sleep cycles, Sleep time &  \checkmark & O, M  & T, M & A & N, R & ML, UD & T, E  \\  

[Health Monitor] BioHarness (zephyranywhere.com) & \catDxxWearable & GPS, ECG, Heart rate  & Breathing rate, Posture, Activity level, Peak Acceleration, Speed, Distance &  \checkmark & O, M, W, D & T, M & RT, A & N, R & ML, UD & E  \\  

[Remote Health Monitor] BodyGuardian (preventice.com) & \catDxxWearable & ECG, Biometric Sensors & Health report &  \checkmark & O, M, W & T, M & RT, A & N, R & ML, UD & T, E  \\  

[Smart Ring] Electricfoxy (electricfoxy.com) & \catDxxWearable & 3-Axis accelerometer, Heart rate, GPS & Heart condition, Calories Burned &  \checkmark & O M & T, M & RT & N & ML, UD & T, S, E  \\  

[Health-Fitness Tracker] Fitbit (fitbit.com) & \catDxxWearable & 3-Axis accelerometer & Steps, Distance, Calories Burned,  Floors Climbed,   Sleep Tracking  &  \checkmark & O, M, W & T, M & RT, A & N, R & ML, UD & T, S, E  \\  

[Emergency Helmet] ICEdot (icedot.org) & \catDxxWearable & Users' medication, Users' personal allergies & Location &  \checkmark & M & M & RT, A & N, A & UD & E  \\

[Fitness Tracker] Lark (lark.com) & \catDxxWearable & Accelerometers, Gravity, Gyroscopes, Rotational vector, Orientation, Magnetometers & Activity recognition, Calories burned &  \checkmark & M & M & RT, A & N, R & UD & T, E  \\

[Sport Watch] Leikr (leikr.com) & \catDxxWearable & GPS, Heart Rate  & Distance, Calories burned,  Speed, Average pace per lap, Lap distance,  Lap calories &  \checkmark & O & T & RT, A & N, R & UD & T, S, E  \\

\hline

\end{tabular}
\end{table*}

%============================= Table 3===================================
%========================================================================

\begin{table*}[t!]

\footnotesize
\renewcommand{\arraystretch}{0.8}
\begin{tabular}{
 p{2.5cm} 
 c 
 m{2.8cm}  
 m{2.8cm}
 c
 c
 c
 c
 c
 c
 c 
 } 
%  \hline 
%  \multicolumn{3}{|c|}{OneTwoThree} \\
%  \\
 \hline
 
  \begin{minipage}[b]{2.0cm}[Project Type] Project Name (Web Link) \end{minipage}      &          %--(1)
 \begin{sideways}Category \end{sideways}   &   %--(2)
 \begin{minipage}[b]{2.0cm}Primary Context \end{minipage}  &                                         
 \begin{minipage}[b]{2.2cm}Secondary Context \end{minipage}  &
 \begin{sideways}\begin{minipage}[b]{1.4cm}Visual \\Presentation\end{minipage} \end{sideways} & %--(7)
 \begin{sideways}\begin{minipage}[b]{1.4cm}Presentation Channel\end{minipage} \end{sideways} & %--(7)
 \begin{sideways}\begin{minipage}[b]{1.4cm}User \\Interaction\end{minipage} \end{sideways} & %--(7)
 \begin{sideways}\begin{minipage}[b]{1.4cm}Real-Time \\ Archival \end{minipage} \end{sideways} &  %--(8)
 \begin{sideways}\begin{minipage}[b]{1.4cm}Notification \\ Mechanism \end{minipage} \end{sideways} &   \begin{sideways}\begin{minipage}[b]{1.4cm}Learning Ability \end{minipage} \end{sideways} &  %--(8)
\begin{sideways}\begin{minipage}[b]{1.4cm}Notification Execution \end{minipage} \end{sideways}   %--(8)
% \begin{sideways}\begin{minipage}[b]{1.6cm}User \\Involvement \end{minipage} \end{sideways}    %--(12) 
 \\  
 \hline \hline 
% \vspace{0.3cm}  
(1)      & (2)              & (3)      & (4)      &(5)            & (6)       & (7)      & (8)    & (9)                            &  (10)   &  (11)                 \\

[Activity Tracker] Lumoback (lumobodytech.com) & \catDxxWearable & 3-Axis Accelerometer & Posture  steps, Distance travelled, Activity recognition, Calories burned &  \checkmark & O, M, D & T, M & RT, A & N, R & UD & T, E  \\  

[Baby Monitor] Mimo  (mimobaby.com) & \catDxxWearable & 3-Axis Accelerometer, Audio, Respiration  & Baby sleep status, Respiration patterns,  Baby's body position &  \checkmark & M, W & T, M & RT, A & N & UD & E  \\  

[Health Monitor] MyBasis (mybasis.com) & \catDxxWearable & Heart rate, Galvanic skin response, Skin temperature, 3-Axis Accelerometer & Activity, Health, Calories &  \checkmark & O, M & T, M & RT, A & N, R & ML & E  \\  

[Medical Jacket] MyTJacket (mytjacket.com) & \catDxxWearable & Pressure & Activity level &  \checkmark & M, W & T, M & RT, A & N, A & ML, UD & E  \\  

[Security Authenticator] Nymi (nymi.com) & \catDxxWearable & Heart activity & Personal Identity &  \checkmark & O & T, M & - & N & - & E  \\  

[Sport Goggles] Oakley  (oakley.com) & \catDxxWearable & GPS, 3-Axis Accelerometer
3-Axis Gyroscope, 3-Axis Magnetometer, Temperature, Barometric Pressure & Speed, Track friends, Navigation maps, Jump Analytic &  \checkmark & O, M, W & T, M & RT, A & N, R, A & UD, ML & T, S, E  \\  

[Sports Helmet] TheShockBox (theshockbox.com) & \catDxxWearable & Accelerometer, Rotation, Pressure &  Hit direction,  Force estimation, Hit count &  \checkmark & M, W & T, M & RT, A & N, R & UD & E  \\  

[Sport Assistant] Zepp (zepp.com) & \catDxxWearable & Dual accelerometers
3-Axis Gyroscope & 3D swing, Club speed, Swing plane, Tempo, Backswing position, Hip rotation &  \checkmark & M & T, M & RT, A & N,R & ML,UD & E  \\

[Indoor Air Quality Monitor] Alima (getalima.com) & \catExxSmartHome & Volatile organic compounds, CO$_{2}$, CO,  Temperature, Humidity, Accelerometer,  & Indoor air quality prediction &  \checkmark & O, M, W & M & RT, A & N & ML & E  \\  

[Smart Locator] BiKN (bikn.com) & \catExxSmartHome & Beacon signal strength & Distance, Geo-fencing &  \checkmark & O,M & T, M & RT, A & N & UD & S  \\  

[Family Connections] Good Night Lamp (goodnightlamp.com) & \catExxSmartHome & $ \times $ & $ \times $ & $ \times $ & O & T & RT, A & A & $ \times $ & $ \times $  \\  

[Light Bulb] Hue  Bulb (meethue.com) & \catExxSmartHome & $ \times $ & $ \times $ &  \checkmark & M & M & $ \times $ & A & UD & $ \times $  \\  

[Door Lock] Lockitron (lockitron.com) & \catExxSmartHome & GPS, Person ID & Identify family and friends &  \checkmark & O, M & M & - & A & - & T, S, E  \\  

[Smart Thermostat] Nest (nest.com) & \catExxSmartHome & Temperature & Efficient heating schedule, Heat up and cool down time calculation &  \checkmark & O, M & T, M & RT & A & ML & E  \\  

[Smart Home] Ninja Blocks (ninjablocks.com) & \catExxSmartHome & Motion, Moisture, Temperature,  Light, Humidity, Presence [extendible] & Energy usage, Indoor localization &  \checkmark & M & V, M & RT & N, R, A & UD & T, S, E  \\

[Weather Station] Netatmo (netatmo.com) & \catExxSmartHome & Temperature, Humidity, Air quality, CO$_{2}$, Sound, Pressure & Weather prediction &  \checkmark & M, W & M & RT & N & UD & E  \\  

[Smart Scale] Withings (withings.com) & \catExxSmartHome & Weight, Body composition, Heart rate,  Temperature, CO$_{2}$ & Body Mass Index, Air quality, Automatic user recognition &  \checkmark & M, W & M & RT, A & N, R, A & UD & $ \times $  \\  

[Smart Home] SmartThings (smartthings.com) & \catExxSmartHome & Motion, Moisture, Temperature,  Light, Humidity, Presence [extendible] & Energy usage, Indoor localization &  \checkmark & M & V, M & RT & N, R, A & UD & T, S, E  \\

[Thermostat] Tado (tado.com) & \catExxSmartHome & Temperature, GPS, Weather forecast & Efficient heating schedule , User location prediction &  \checkmark & M & M & RT, A & N, A & ML, UD & T, S  \\

[Smart Cooking] Twine (supermechanical.com) & \catExxSmartHome & Moisture, Magnetism, Temperature, Vibration,  Orientation & Recommendation to cook meat &  \checkmark & M, W & M & RT, A & N & UD & T, S, E  \\

\hline

\end{tabular}
\end{table*}

%============================= Table 4===================================
%========================================================================

\begin{table*}[t!]

\footnotesize
\renewcommand{\arraystretch}{0.8}
\begin{tabular}{
 p{2.5cm} 
 c 
 m{2.8cm}  
 m{2.4cm}
 c
 c
 c
 c
 c
 c
 c 
 } 
%  \hline 
%  \multicolumn{3}{|c|}{OneTwoThree} \\
%  \\
 \hline
 
  \begin{minipage}[b]{2.0cm}[Project Type] Project Name (Web Link) \end{minipage}      &          %--(1)
 \begin{sideways}Category \end{sideways}   &   %--(2)
 \begin{minipage}[b]{2.0cm}Primary Context \end{minipage}  &                                         
 \begin{minipage}[b]{2.2cm}Secondary Context \end{minipage}  &
 \begin{sideways}\begin{minipage}[b]{1.4cm}Visual \\Presentation\end{minipage} \end{sideways} & %--(7)
 \begin{sideways}\begin{minipage}[b]{1.4cm}Presentation Channel\end{minipage} \end{sideways} & %--(7)
 \begin{sideways}\begin{minipage}[b]{1.4cm}User \\Interaction\end{minipage} \end{sideways} & %--(7)
 \begin{sideways}\begin{minipage}[b]{1.4cm}Real-Time \\ Archival \end{minipage} \end{sideways} &  %--(8)
 \begin{sideways}\begin{minipage}[b]{1.4cm}Notification \\ Mechanism \end{minipage} \end{sideways} &   \begin{sideways}\begin{minipage}[b]{1.4cm}Learning Ability \end{minipage} \end{sideways} &  %--(8)
\begin{sideways}\begin{minipage}[b]{1.4cm}Notification Execution \end{minipage} \end{sideways}   %--(8)

 \\  
 \hline \hline 
% \vspace{0.3cm}  
(1)      & (2)              & (3)      & (4)      &(5)            & (6)       & (7)      & (8)    & (9)                            &  (10)   &  (11)                 \\

[Personal Assistant] Ubi (theubi.com) & \catExxSmartHome & Temperature, light, humidity, pressure  & - &  \checkmark & O, M, W & V & RT, A & N, R, A & ML, UD & T, S, E  \\  

[Power Plug] WeMo Switch (belkin.com) & \catExxSmartHome & Temperature, energy consumption & Estimate Cost &  \checkmark & M & T, M & RT, A & N, A & UD & T,E  \\  

[Family Connections] WhereDial (wheredial.com) & \catExxSmartHome & GPS & location (e.g. pub, work, home) & $ \times $ & O & T & RT & N &  $ \times $ & E  \\  

[Dog Activity Monitoring] Whistle (whistle.com) & \catExxSmartHome & Accelerometer,  location, person & Daily Activity Report (play time, rest time), Medical Recommendations, Excers &  \checkmark & M & M & RT, A & N, R & ML & S  \\

%\catAxxSmartCity \catBxxEnterprise \catCxxEnvironment \catExxSmartHome \catDxxWearable
\hline

\end{tabular}
\end{table*}

  \section{Lessons Learned, Opportunities and Challenges}
  \label{sec:Lessons_Learned}
This section presents some major lessons we learnt during the IoT product review.
 
 \subsection{Trends and Opportunities}
    
According to our survey on the IoT product marketplace, it is evident that the types of primary context information collected through sensors are mostly limited. However, the ways such collected data is been processed  varied significantly based on the application and the required functionalities that the IoT product plan to offer. Therefore, it is important to understand that, in IoT, same data can be used to derive different insights in different domain. In combine, the IoT solutions have used around 30-40 different types of sensors to measure different parameters. The ability to derive different insights using same set of data validates the importance of sensing as a service model \cite{ZMP008}, which envisions to create a data market that buys and sells data. 

Most of the IoT solutions have used some kind of context presentation technique that summarizes and converts the data into a easily understandable format. It is also important to note that, despite the advances in human computer interaction, most of the IoT solutions have only employed traditional computer screen-based technique. Only few IoT solutions really allow voice or object-based direct communications. However, most of the wearable solutions use touch as a common interaction technique. We also see a trend of smart home products also increasingly use touch-based interactions. Hands free voice or gesture based user interaction will help consumers to seamlessly integrate IoT products into their lives. At least, smart watches and glasses may help to reduce the distraction that smart phones may create when interacting with IoT products.

Most of the IoT products ends their services after releasing notification to the consumers. Users will need to perform the appropriate actuation tasks manually. Lack of standards in machine to machine (M2M) communication seems to play a significant role in this matter. We will discuss this issue in Section \ref{sec:Interoperability}. Finally, it is important to note that increasing number of IoT products use data analytic and reasoning in order to embed more intelligence to their products. As a result, there is a need for domain independent, easy to use (e.g. drag and drop configuration without any program coding) analytical frameworks with different characteristics where some may  effectively  perform on the cloud and the others may work efficiently in resource constrained devices. One solution in this space is Microsoft Azure Machine. Learning\footnote{http://azure.microsoft.com/en-us/services/machine-learning/}. Another generic framework is \textit{Wit}. \textit{Wit} (Wit.ai) is a  natural language processing API for the IoT which allows developers to easily and quickly  add natural language processing functionality to their IoT solutions.
 
 It is important to note that most of the IoT solutions consider families or group of people as a whole, not as individuals. Therefore, most of the IoT solutions are unable to individually and separately identify father, mother or child living in a given house. For example, the temperature that individual family members would like to have can be different. However, most of the modern thermostat only consider context information such as past behaviour, time of the day, presence of a user, and so on. However, it cannot handle individual preferences of the family members. Therefore, embedding such capabilities to the IoT products would be a critical requirement to be successful in  future IoT marketplace.
 
 In order to support and encourage the adoption of IoT solution among consumers, it is important to make sure that the usage of products allows to recover the cost of product purchase within a reasonable time period. For example, the \textit{Nest} thermostat promises that consumers can recover its costs through reducing the energy bill. Auto-Schedule feature in \textit{Nest} makes it easy to create an energy efficient schedule that  help the users to save up to 20\% on  heating and cooling bills.

  \subsection{Product Prototyping}
\begin{figure}[!b]
     \centering
    % \vspace{-0.43cm}
     \includegraphics[scale=0.57]{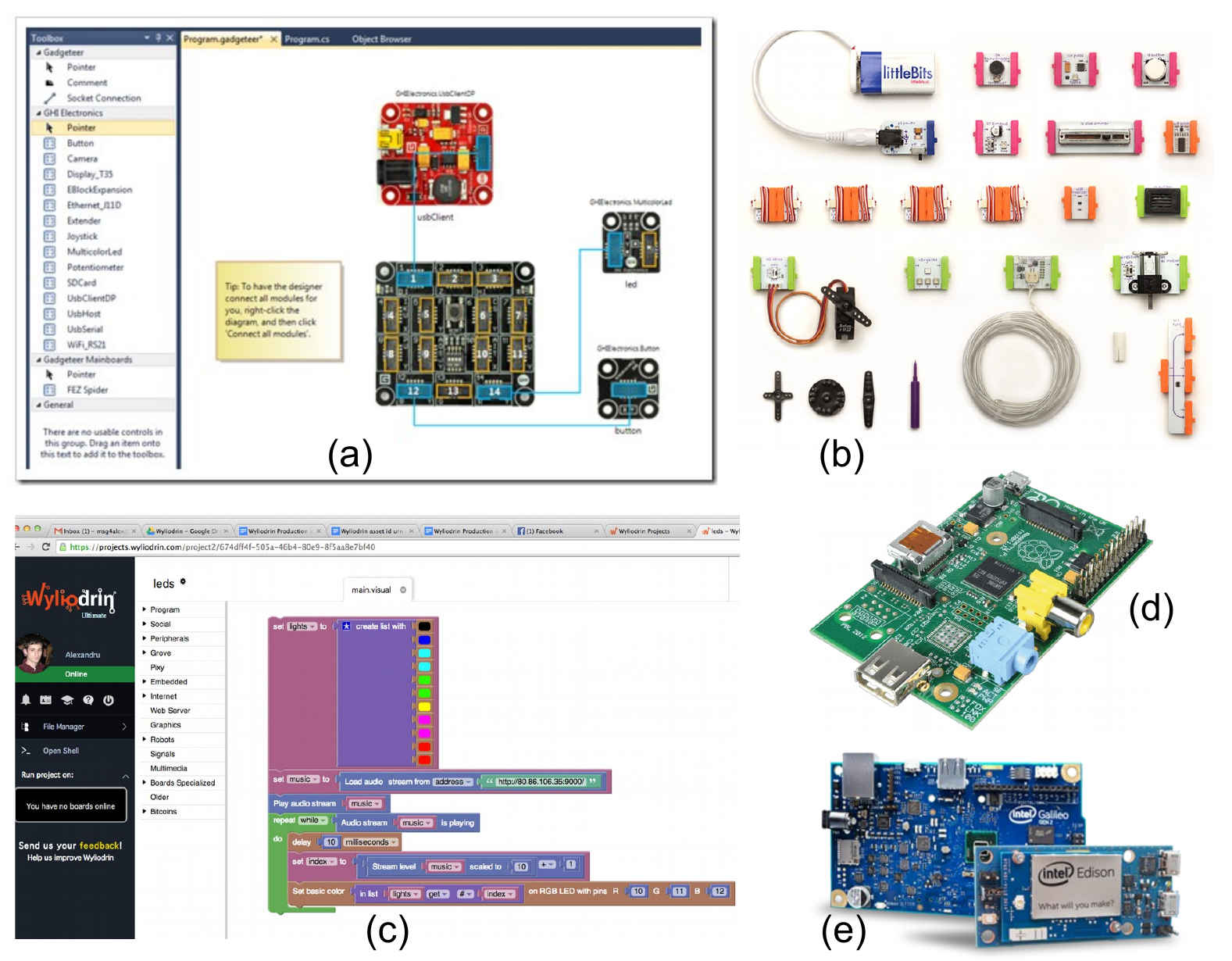}
    %\vspace{-0.33cm}	
     \caption{(a) Microsoft Visual Studio IDE that allows to visually wire \textit{.NET Gadgeteer} hardware components. The IDE automatically generated the code skeletons to make the prototyping much easier and faster, (b) Hardware sensors and actuators of \textit{LittleBits} (littlebits.cc) platform, (c) Wyliodrin web-based IDE that allows to program variety of different platforms including \textit{Arduino} (arduino.cc) and  \textit{Raspberry Pi} (www.raspberrypi.org) by visually drag and drop programming components, (d) a \textit{Raspberry Pi} (www.raspberrypi.org), (e) \textit{Intel Edison} board.}
     \label{Figure:Prototyping_Tools}	
    \vspace{-0.4cm}	
    \end{figure}

  There are number of do-it-yourself (DIY)  prototyping platforms available that allows to create IoT prototypes quickly and easily. Specially these platforms are cheaper and modular in nature. They allow anyone with a new idea to test their initial thoughts with very limited budget, resources, and more importantly less time.  \textit{Arduino} (arduino.cc) (including variations such as \textit{Libelium} (libelium.com)), \textit{.NET Gargeteer} (netmf.com/gadgeteer), \textit{LittleBits} (littlebits.cc) are some well known prototyping platforms. Most of these products are open source in nature. More importantly over the last few years, they have become more interoperable which allows product designers to combine different prototyping platforms together. The programming mechanisms  use to program these modules can be  varied (e.g. C, C++, C\#, Java, Javascript, etc.). Some platforms provide  easy and intuitive ways to write program such as mashing-ups and wirings as shown in Figure \ref{Figure:Prototyping_Tools}.

  There are  small computer systems been developed to support IoT prototyping. For example \textit{Raspberry Pi} (www.raspberrypi.org) is a such product.  \textit{Raspberry Pi} is a credit card-sized single-board computer developed in the UK by the \textit{Raspberry Pi} Foundation with the intention of promoting the teaching of basic computer science in schools. However, more recently, \textit{Raspberry Pi}s are heavily used in IoT product prototype development. For example, IoT products such as \textit{NinjaBlocks} (ninjablocks.com) has used \textit{Raspberry Pi}s in their production officially. Further, most of the platforms such as \textit{Ardunio} can successfully work with \textit{Raspberry Pi} Computers. Recently, Intel has also produced a small computer (e.g. Intel Galileo and Intel Edison boards) competitive to \textit{Raspberry Pi} which runs both windows and Linux. The Intel Edison is a tiny computer offered by \textit{Intel} as a development system for wearable devices.
 
 Programming IDE tools such \textit{Microsoft Visual Studio} provides significant support for IoT program development by facilitating visual  wiring,  mash ups and automated code generation. Such ease of programming and prototyping abilities have attracted significant attention from hobbyist, researcher, and even from school children.
 
 These modular based prototyping tools allow to build and test context-aware functionalities efficiently and effectively. Most of these platforms offer large number of sensing modules that allow to collect data from different types sensors. As we mentioned earlier such data can be considered as primary context. Therefore, such primary context can be combined together to generate secondary context information. However, in most of the prototyping platforms, secondary context discovery needs to be done manually or using IF-ELSE statements. However, it would be much useful to develop a standard framework with modularity in mind to address this issue. These modules need to be defined in a standard form despite their differences in real implementations. Further, such context discovery modules should be able to combine together to discover more advance context information \cite{IA001}. We further explain how such framework should work in real world in Section \ref{sec:Resources}.

  \subsection{Interoperability on Product and Services}
\label{sec:Interoperability}
Interoperability is a critical factor to be successful in IoT domain. Consumers typically do not want to stick into one single manufacturer or service provider. They always go for their preferences and for the factor which are more important to them such as cost, look and feel, customer service, functionality and so on. Interoperability among different IoT products and solutions allows consumers to move from one product to another or combine multiple products and services to build their smart environments as they like in a customize fashion. Further, interoperability \cite{IA07} is also important to eliminate market domination of large companies that increase the entry barriers for the small IoT product and service providers.

In IoT market place, interoperability is mainly achieved using three methods: 1) partnerships among product and service developers, 2) open and close standards, and 3) adaptors and mediator services. We have seen that major industrial players in the IoT marketplace stablish strategic partnerships with each other in order to enable interoperability among their product and services. However, this is not a scalable strategy to widely enable interoperability among IoT devices. Similarly, large corporations such as Apple (e.g. \textit{HomeKit}\footnote{developer.apple.com/homekit}, \textit{HealthKit}\footnote{developer.apple.com/healthkit}) and \textit{Google} (e.g. Fit\footnote{developers.google.com/fit}) are also attempting to build their own standards and interoperability certifications. This kind of interoperability may lead to corporate domination of IoT marketplace which could also hinder the innovation by small, medium, and start-up companies.

     \begin{figure}[!t]
      \centering
     % \vspace{-0.43cm}
      \includegraphics[scale=0.48]{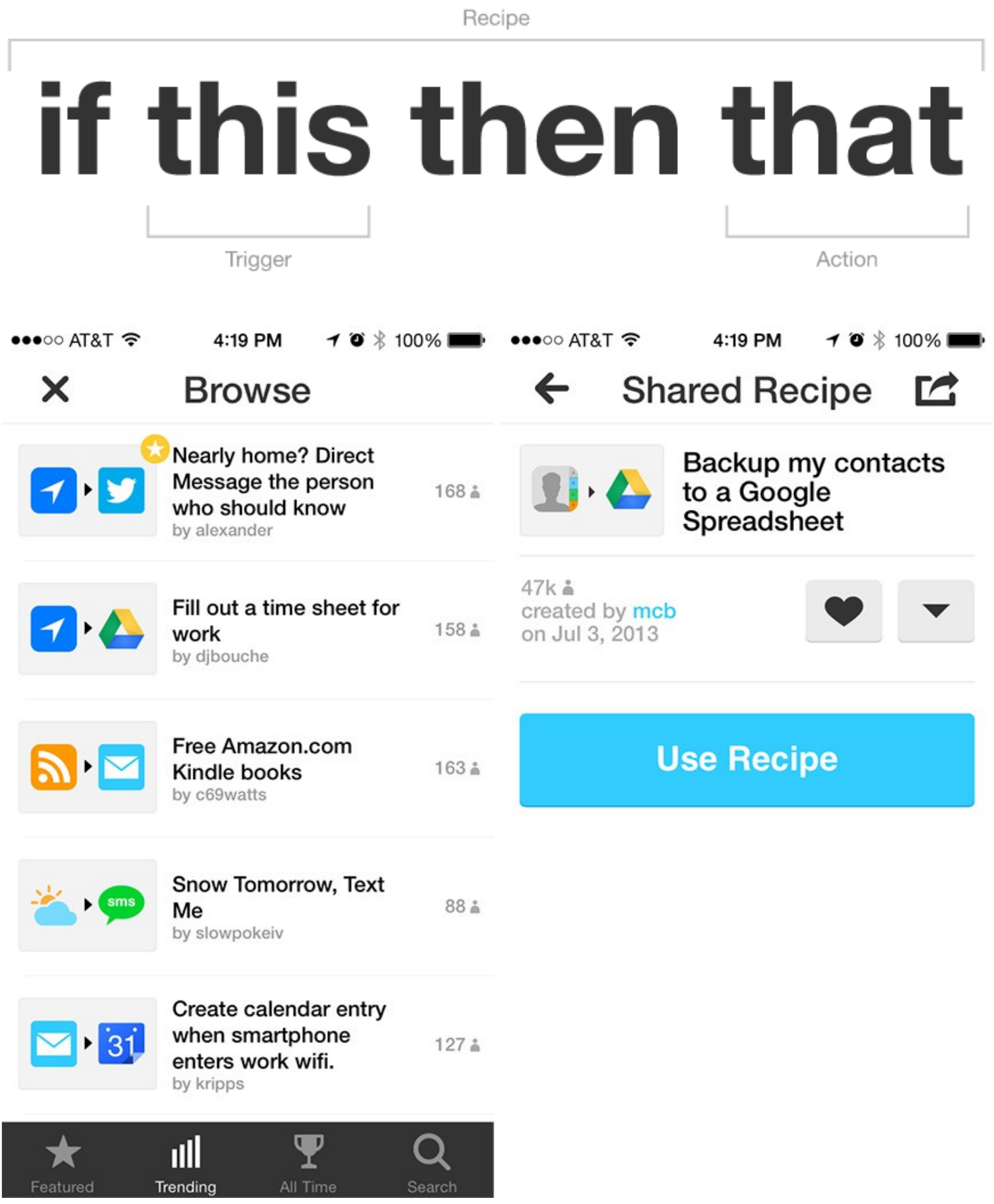}
     %\vspace{-0.33cm}	
      \caption{(a) shows how a recipe is structured using conditional statements and actions. (b) shows how recipes are built combining different triggers, actions, and channels.}
      \label{Figure:Interoperability}	
     %\vspace{-0.4cm}	
     \end{figure}

To address the interoperability, there are some alliance have been initiated. For example \textit{AllSeen Alliance} (allseenalliance.org) has been created to promote some kind of interoperability among IoT consumer brands. \textit{AllSeen} has developed a standard software platform called \textit{AllJoyn}. \textit{AllJoyn} is a system that allows devices to advertise and share their abilities with other devices around them. A simple example would be a motion sensor letting a light bulb know no one is in the room it is lighting. This is the ideal approach the interoperability among IoT products. However, security \cite{IA03} and privacy in this framework need to be strengthen to avoid using interoperability features to attack IoT products by hackers or evil parties.

Another approach to enable interoperability among different IoT solutions is through adapter services. For example, \textit{IFTTT} (ifttt.com), If This Then That, is a web based service that allows users to create powerful connections, chains of simple conditional statements. One simple statement is illustrated in Figure \ref{Figure:Interoperability}. Channels are the basic building blocks of \textit{IFTTT}. Each Channel has its own \textit{Triggers} and \textit{Actions}. Some example Channels could be \textit{Facebook}, \textit{Twitter}, weather,\textit{ Android Wear}, and so on. Channel could be both hardware or software. Service providers and product manufactures need to register their services with \textit{IFTTT} once. After that anyone interested ca  use that product or service as a channel to compose any recipe. Example list of channels are listed here: ifttt.com/channels.  Personal recipes are combinations of a \textit{Trigger} and an \textit{Action} from  active Channels. Example recipes are shown in Figure \ref{Figure:Interoperability}. For example, first recipe is defined to send a twitter message to a family member when the user reaches home. This kind of recipe can be used to offload responsibility from a child so the system automatically act on behalf of the child and sent a tweet to their parents. Context-aware recommendation can also help users to quickly configure channels in \textit{IFTTT}. Context could be location, time, family members around, IoT products located near by and so on. Context-aware recommendation \cite{IA003} can also be done by analysing similar users with similar smart environments.

    \subsection{Resources and Energy Management}
    \label{sec:Resources}
    
  Most popular approach of energy management in IoT is through smart plugs.  \textit{Plugwise} (shop.plugwise.com), \textit{Thinkecoinc} (shop.thinkecoinc.com), \textit{Belkin} (www.belkin.com) provide similar functionalities and services where they capture energy consumption using smart plugs. These solutions analyse data in many different ways and presented the context information to the users using variety of different charts and graphs. These plugs can also be used to home automation as they can be switched ON and OFF remotely or conditionally. For example, a condition would be temporal (i.e. time-aware behaviour) or spatial (i.e. location-aware behaviour). 
  
   There aren't any IoT solutions that focus on planning or deployment stages of smart environments. Analyse energy consumption is important in both industrial large scale deployments (e.g. waste management solutions discussed in \cite{ZMP008}) and in consumer based smart home and office deployments.   Lets consider a smart home office planing and deployment scenario. At the moment, IoT marketplace is flooded with large number of IoT smart products that offer different functionalities. However, there aren't any method for consumers to measure or compare the benefits these products may offer and the associated costs such as  cost of purchase, installation and maintains. Further, it is very hard to understand which solutions can work together and complement each other and which work standalone. 
   
   It is also difficult to understand where to install certain smart products and how many products are required to cover a certain area. (e.g. what are the ideal locations to install micro-climate sensors within a building which enable to accurately identify the micro-climate behaviour).  Another issue would be to determine the coverage of a product. For example, how many motion sensors are required for a given home or office. Currently, to best of our knowledge, there is no such tool that  can be used to achieve above planning and installation tasks. As we mentioned before, consumers are always eager to know the costs and benefits of a products. Therefore, it is important to facilitate some tools that can demonstrate cost benefit analysis (e.g. purchase cost, maintenance cost such as energy, energy saving and so on.). Context information will play a significant role in this kind of tools where consumers may need to input the budget, size of the building, their priorities and expectations. The tool will need to make recommendations to the consumers on which product to buy based on the product's technical specification and other consumers' reviews and comments.

   The planing and installation becomes much more critical in industrial settings. Let considers the agricultural sensing scenario, the \textit{Phenonet} project, presented in \cite{ZMC008}. \textit{Phenonet} describes    the network of sensors collecting information over a    field of experimental crops. Researchers at the \textit{High    Resolution Plant Phenomics Centre} \cite{P585} needs to  monitor plant growth and    performance information under different climate conditions over time.
   
   It would be very valuable to have a tool that can help planning large scale sensor deployments. For example, energy predictive models will help the users to decide what kind of energy sources to be used and what kind of battery size to be used in each scenario. The amount of sensor nodes require to cover a curtain geographical area should be able to accurately predicted based on the context information using such tool. For example, in the agricultural sensing scenario, sensors  deployments are  planned by agricultural scientist who have little knowledge on  electronic, communication, or energy consumption. Therefore, it is useful to have a user friendly tool that enables them to plot and visualise a large scale sensor deployment in virtual setting before getting into real world deployments. Perera et al. \cite{ZMC008} have present the agriculture scenario in detail.

    Context information plays a critical role in sensor configuration in large scale sensor deployments in IoT. The objective of collecting sensor data is to understand the environment better by fusing and reasoning  them. In order to accomplish this task, sensor data needs to be collected in a timely and location-sensitive manner. Each sensor needs to be configured by considering context information. Let us consider a scenario related to smart agriculture to understand why context matters in sensor configuration. \textit{Severe frosts and heat events can have a devastating effect on crops. Flowering time is critical for cereal crops and a frost event could damage the flowering mechanism of the plant. However, the ideal sampling rate could vary depending on both the season of the year and the time of day. For example, a higher sampling rate is necessary during the winter and the night. In contrast, lower sampling would be sufficient during summer and daytime. On the other hand, some reasoning approaches may require multiple sensor data readings. For example, a frost event can be detected by fusing air temperature, soil temperature, and humidity data. However, if the air temperature sensor stops sensing due to a malfunction, there is no value in sensing humidity, because frost events cannot be detected without temperature. In such circumstances, configuring the humidity sensor to sleep is ideal until the temperature sensor is replaced and starts sensing again}. Such intelligent (re-)configuration can save energy by eliminating  ineffectual  sensing and network communication.
    
   An ideal tool should be able to simulate different types of user scenarios virtually before the real world deployments begin.   Once deployed, another set of tools are required to advice and recommend, scientists and non-technical users, on configuring sensor parameters. Configuring sensors in a optimal fashion would lead to longer operation time while maintaining required accuracy. It is important to develop the tools in a modular and standard fashion so the manufacturers of each IoT solution can add their products into a library of product which enables consumers to easily select (may be drag and drop and visualize) the product they prefer for visualization purposes. Further, such tools will need to be able to combine different compatible products together autonomously based on context information such as budget, user preferences, and location information so the users will be offered different combinations to select from.

Resource  management is also a critical task that need to be done optimally in IoT domain. Previously, we discussed how data may transferred over the network as well as through different types of data processing devices in Figure \ref{Figure:Context_aware_Big_Picture}. It is hard to determine the optimal location\footnote{the device that is responsible for processing data} to process data. Therefore, it is  ideal to have a tool that is capable of evaluating a given software component\footnote{A self contained algorithm that may take primary context information as inputs and outputs secondary context information using any kind of data reasoning technique \cite{ZMP007}.} against a given   computational network architecture and deciding which location is optimal to conduct any kind of reasoning  based on user preferences, resource availability, context information availability, network communication availability and so on.

    \subsection{Privacy and Data Analytic}
    
 IoT marketplace is mainly composed with three parties, namely: device manufacturers, IoT cloud services and platform providers, and third party application developers [15]. All these parities need to consider privacy as a serious requirement and a challenge. In this section, we present some advice on preserving user privacy in IoT domain.     
       
\textit{Device Manufacturers:} Device manufactures must embed privacy preserving techniques into their devices. Specially, manufactures must implement secure storage, data deletion, and control access mechanisms at the firmware level. Manufactures must also inform consumers about the type of data that are collected by the devices. Moreover, they must also explain what kind of data processing will be employed and how and when data would be extracted out of the devices. Next, the manufactures must also provide the necessary control for the consumers to disable any hardware components. For example, in an IoT security solution, consumers may prefer to disable the outside CCTV cameras when inside the home. However, consumers will prefer to keep both inside and outside cameras active when they leave the premises. Moreover, devices manufactures may also need to provide programming interface for third party developers to acquire data from the devices. 
 
\textit{IoT Cloud Services and Platform Providers:} It is likely that most of the IoT solutions will have a cloud based service that is responsible for proving advance data analysis support for the local software platforms. It is very critical that such cloud providers use common standards, so that the consumers have a choice to decide which provider to use. Users must be able to seamlessly delete and move data from one provider to another over time. Such a possibility can only be achieved by following a common set of interfaces and data formats. Most of the cloud services will also use local software and hardware gateways such as mobile phones that act as intermediary controllers. Such devices can be used to encrypt data locally to improved security and to process and filter data locally to reduce the amount of data send to the cloud. Such methods will reduce the possibility of user privacy violation that can occur during the data transmission.

\textit{Third Part Application Developers:} Application developers have the responsibility to certify their apps to ensure that they do not contain any malware. Moreover, it is the developers' responsibility to ensure that they present clear and accurate information to the users to acquire explicit user consent. Some critical information are: (1) the task that the app performs, (2) the required data to accomplish the tasks, (3) hardware and software sensors employed, (4) kind of aggregation and data analysis techniques that the app will employ, (5) kind of knowledge that the app will derive by data processing. 
Users need to be presented with a list of features that the application provides, and the authorization that the user needs to give to activate each of those features. The control must be given to the user to decide which feature they want to activate. Moreover, in the IoT, acquiring user consent should be a continuous and ongoing process. Consequently, the application developers must continuously allow the users to withdraw, grant, or change their consent. Moreover, users must be given full access to the data collected by the IoT devices.

   \subsection{Central Hubs}
\begin{figure}[!h]
       \centering
      % \vspace{-0.43cm}
       \includegraphics[scale=0.48]{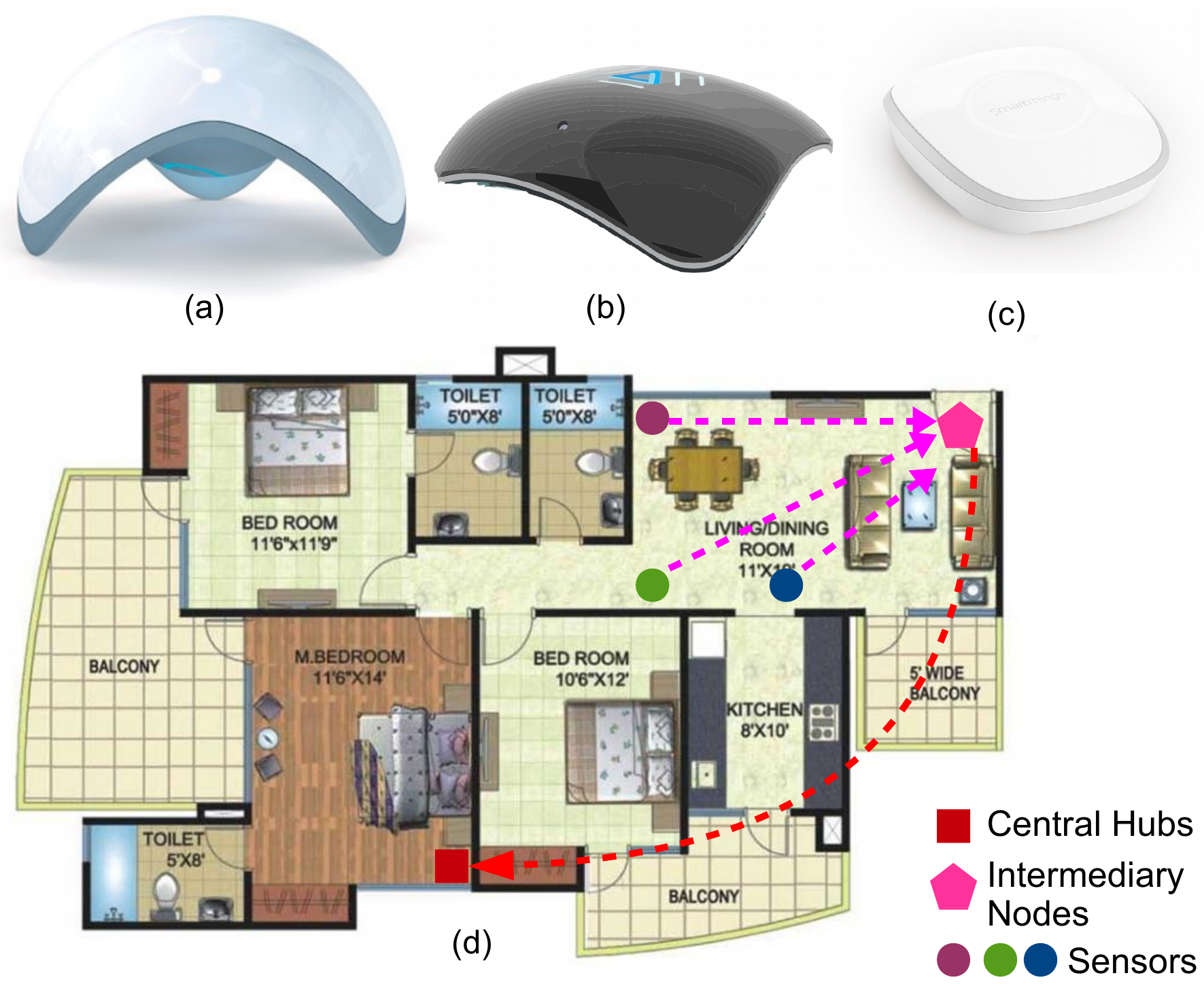}
      %\vspace{-0.33cm}	
       \caption{Centralised Hubs are category of devices heavily used in IoT solution.  (a) \textit{Ninja sphere} (b) \textit{ALYT} Hub powered by \textit{Andorid} (c) \textit{Samsung}'s \textit{SmartThing} Hub (d) Sensors and other components are connected to a centralised hub. These hubs are typically connected to permanent power sources and comprises comparatively high computational capabilities.}
       \label{Figure:Home_Hubs}	
      %\vspace{-0.4cm}	
      \end{figure}

Central hubs are commonly used in IoT solutions. A typical IoT solution may comprises a  number of different components. For example, an IoT solution may have sensors, actuators, processing and communication devices. Due to the nature, sensors and actuators may need to deploy in certain location manner (e.g. door sensor must mount on the door). As a result such sensors and actuators need to be small in size. Due to miniature size, it is not possible to enrich them with significant computational capacity. Similarly, most of the time these sensors and actuators would be battery powers (i.e. without having connected to permanent power sources). Therefore, energy management within those sensors and actuators is very critical. As a result, such smaller devices cannot perform significant data processing tasks. On the other hand, these individual devices have only limited knowledge about a given context. For example, a door sensors may only know about the current status of the door. The knowledge that can be derived from such limited amount of data is very constrained. In order to comprehensively understand a given situation, context data from number of sensors and actuators need to be collected, processed, and analysed. To address this issue, most of the IoT solutions have been used a central hubs (sometimes called `home hub') or similar solutions as shown in Figure \ref{Figure:Home_Hubs}.

Typically, central hubs are larger in size compared to sensors and actuators. Further, they are capable of communicating using multiple wireless protocols such as  WiFi, WiFi-direct, Bluetooth ZigBee, Z-wave, and so on. They are also capable of storing data for a significant time period. Typically, only one central hub is required for a large area (e.g. house). These hubs may perform data processing and reasoning tasks (e.g. triggering IF-THEN rules). Further these hubs are typically connected to the cloud services. Dispite the differences in , in high-level, all of these hubs allows to add functionalities over time (i.e. extend the functionalities they may offer), through installing new applications. An app could be a IF-ELSE procedure that explain a certain contextual behaviour as illustrated in Figure \ref{Figure:Event_Trigger}.

The problem in this approach is that each IoT solution designers are eager to design their own centralized hub. Such design approach significantly reduces the interoperability among different products and services in the IoT marketplace. These hubs are tend to use custom firmware and software framework stacks. Unlike operating systems, they are mostly designed to run under specific hardware platforms and configurations. As a result, it makes harder for other IoT solutions to use or utilize other centralized hubs in the marketplace. Centralized hubs typically does not have any user interface. They are controlled and managed using smart phones, tablets, or computers.

In order to stimulate the adoption of IoT solution among consumers, it is important to design a common software platform using common set of standard. The current mobile app market is an ideal model for IoT domains as well where users may install different applications in order to enhance their existing IoT products. Verification is required to check whether the required hardware devices is available to support the intended software application. This is similar to the some mobile app stores validate the phone specification before pushing  the each app to a smart phone.  In comparison to mobile phone domain, IoT domain is slightly complex where  hardware also play a significant role. A one possible solution is to use hardware adaptors. This means when a IoT product manufacture wants to design a product  that is interoperable with a another  hub in the IoT marketplace,  they need to design a hardware adaptor that may handle the interoperability using two-way conversions. 

Finally, it is also important to highlight the necessity of intermediation nodes that can perform multi-protocol communication, bridging short range protocols, and protocol conversions \cite{Z1041}. For example, sensors that may use Bluetooth and ZigBee which can only communicate very short distance. To accommodate such sensors, intermediary nodes may be required.  The intermediate nodes may install throughout a given location which may use with log range protocols to communicate with the central hub.  The intermediate nodes may use short rage protocols to communicate with sensors and actuators.

      \subsection{Legacy Devices}
Most of the IoT products in the marketplace comes with  own hardware components and software stacks. However, we have increasingly seen that  IoT solutions attempt to enrich legacy devices with smart capabilities. One very popular solution is \textit{Nest} (nest.com) thermostat. It has the capability to learn from users over time about their behaviour and preferences and  control the temperature more efficiently and pro-actively. This thermostat can be installed by replacing the existing non-smart traditional thermostats. Everything else connected to the heating systems would work seamlessly.  \textit{ShutterEaze} (shuttereaze.com) is another example for enriching legacy devices. This example is more into home automation. \textit{ShutterEaze} makes it easy for anyone to add remote control functionality and automate their existing interior plantation shutters. No shutters changing is required. 

\begin{figure}[!h]
        \centering
       % \vspace{-0.43cm}
        \includegraphics[scale=0.52]{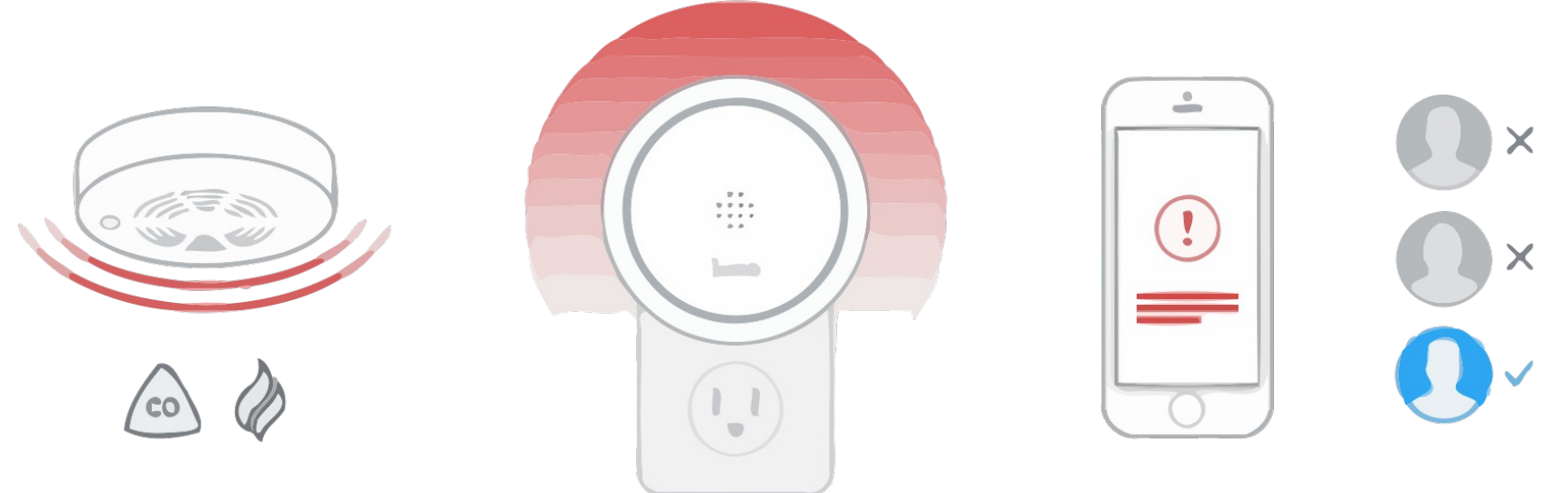}
       %\vspace{-0.33cm}	
        \caption{Enriching smartness to legacy devices: Legacy devices may  monitor fire and smoke. Once these legacy devices detect any abnormalities, they will trigger their alarms and start to make sounds. \textit{Leeo} is designed to listen to such alarm sound. Once \textit{Leeo} detects such sound, it triggers its reaction mechanisms such as sending notification to the users, neighbours, and government authorities such as fire brigade in a predefined order. }
        \label{Figure:Legacy_Device}	
       %\vspace{-0.4cm}	
       \end{figure}

A slightly different example is \textit{Leeo} (leeo.com). As illustrated in Figure \ref{Figure:Legacy_Device}, \textit{Leeo} keeps track of smoke alarms,  carbon monoxide alarms, and the climate in  home. If something is not right, it sends notifications straight to the users phone. It is important to note that, there is no communication between the legacy smoke detection devices / alarms and the \textit{Leeo} device. They are completely two different systems without any dependencies. \textit{Leeo} get triggered by the sound that may produce by other traditional alarms. This is a very good examples to demonstrate  how to embed smartness to our homes without replacing existing legacy systems. More importantly, any kind of replacing cost a significant amount to the consumers. This kind of solutions eliminates such unnecessary and extra costs that may put consumers away from adopting IoT solutions. The lesson we can learn is that if the legacy devices cannot understand the context it operates and act intelligently, the new devices can be incorporated to embed smartness to the overall system where new devices helps to mitigate the weaknesses in the legacy devices.

 \section{Concluding Remarks}
 \label{sec:Conclusions}
 
In this survey, we reviewed significant number of  IoT solutions in the industry marketplace from context-aware computing perspective. We briefly highlighted the evolution of context-aware technologies and how they have become increasingly popular and critical in today's applications. First, we reviewed number of IoT products in order to identify context-aware features they support. At the same time, we also categorized the IoT solutions in the market into five different segments: smart wearable, smart home, smart city, smart environment, and smart enterprise. Finally, we identified and discussed seven major lessons learned and opportunities for future research and development in context-aware computing domain. Our ultimate goal is to build a foundation that helps us to understand what has happened in the IoT marketplace in the past so we can plan for the future more efficiently and effectively.

%We believe further research that addresses innovation opportunities will help to develop more interesting IoT solutions and strengthen exciting solutions in this area in both the industrial and the academic sectors.  
 
% \section{ACKNOWLEDGEMENT}
% \label{sec:ACKNOWLEDGEMENT}
% 
% The authors acknowledge support from SSN TCP, CSIRO, Australia and ICT OpenIoT Project, which is co-funded by the European Commission under the Seventh Framework Program, FP7-ICT-2011-7-287305-Openthe IoT. The authors acknowledge the help and contributions from The Australian  National University.

% are expected be benefited and create value out of IoT solutions and technologies
% Towards this, research efforts are ongoing to address in this domain by both internal research and development deparments as well as external research organization.

% \section{ACKNOWLEDGEMENT}
% \label{sec:ACKNOWLEDGEMENT}
% 
% The authors acknowledge support from SSN TCP, CSIRO, Australia and ICT OpenIoT Project, which is co-funded by the European Commission under the Seventh Framework Program, FP7-ICT-2011-7-287305-OpenIoT. The Author(s) acknowledge help and contributions from The Australian  National University.
 
\ifCLASSOPTIONcaptionsoff
  \newpage
\fi



%\bibliographystyle{IEEEtran}

% \def\IEEEbibitemsep{0pt plus .3pt}
% \bibliography{IEEEabrv,Bibliography}
% \bibliographystyle{IEEEtran}
\bibliography{Bibliography}
\bibliographystyle{IEEEtran}

%\begin{thebibliography}{1}
%\end{thebibliography}

%\begin{biography}[{\includegraphics[width=1in,height=1.25in,clip,keepaspectratio]{mshell}}]{Michael Shell}

\begin{IEEEbiography}[{\includegraphics[width=1in,height=1.25in,clip,keepaspectratio]{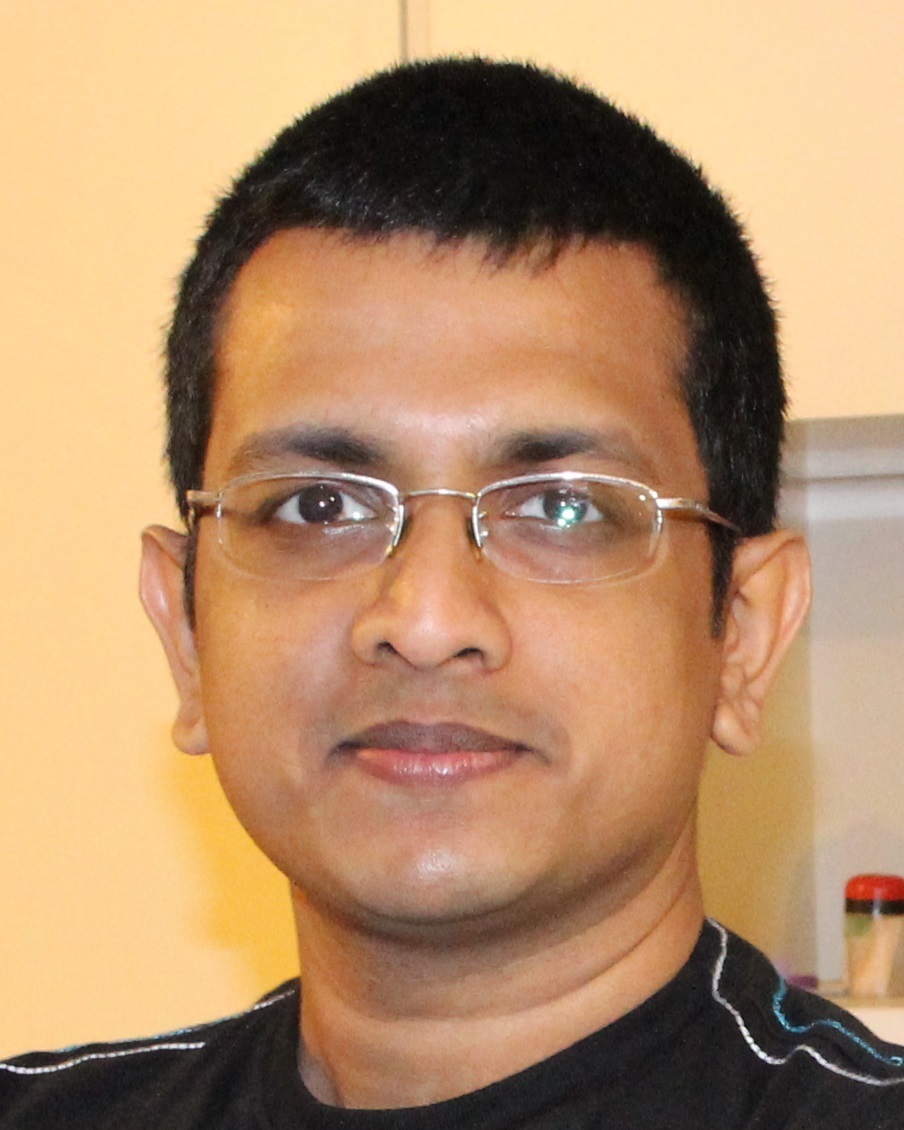}}]{Charith Perera}
received his BSc (Hons) in Computer Science in 2009 from Staffordshire University, Stoke-on-Trent, United Kingdom and MBA in Business Administration in 2012 from University of Wales, Cardiff, United Kingdom and PhD in Computer Science at The Australian National University, Canberra, Australia. He is also worked at Information Engineering Laboratory, ICT Centre, CSIRO and involved in OpenIoT Project (FP7-ICT-2011.1.3) which is co-funded by the European Commission under seventh framework program. He has also contributed into several projects including EPSRC funded HAT project (EP/K039911/1)  His research interests include Internet of Things, Smart Cities, Mobile and Pervasive Computing, Context-awareness, Ubiquitous Computing. He is a member of both IEEE and ACM.
\end{IEEEbiography}

\begin{IEEEbiography}[{\includegraphics[width=1in,height=1.25in,clip,keepaspectratio]{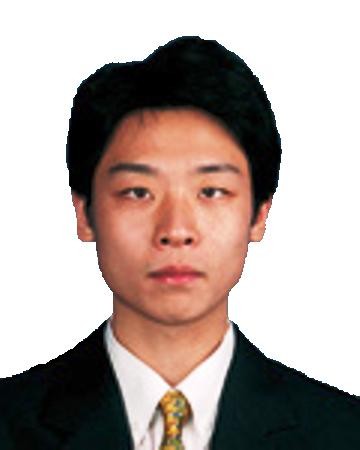}}]{Chi Harold Liu}
is currently a Full Professor with the School of Software, Beijing
Institute of Technology, Beijing, China. He is the Director of the IBM Mainframe Excellence Center (Beijing), the Director of the IBM Big Data Technology Center, and the Director of the National Laboratory of Data Intelligence for China Light Industry. He received the Ph.D. degree from Imperial College London, London, U.K., and the B.Eng. degree from Tsinghua University, Beijing. Before moving to academia, he joined IBM Research, Beijing, as a Staff Researcher and the Project Manager, after working as a Post-Doctoral Researcher with Deutsche Telekom Laboratories, Berlin, Germany, and a Visiting Scholar with the IBM Thomas J. Watson Research Center, Yorktown Heights, NY, USA. His current research interests include the Internet of Things (IoT), big data analytics, mobile computing, and wireless ad hoc, sensor, and mesh networks. He was a recipient of the Distinguished Young Scholar Award in 2013, the IBM First Plateau Invention Achievement Award in 2012, and the IBM First Patent ApplicationAward in 2011, andwas interviewed byEEWeb as the Featured Engineer in 2011. He has authored over 60 prestigious conference and journal papers, and owned over 10 EU/U.S./China patents. He serves as an Editor of the KSII Transactions on Internet and Information Systems and a Book Editor of four books published by Taylor and Francis Group, USA. He served as the General Chair of the IEEE SECON'13 Workshop on IoT Networking and Control, the IEEE WCNC'12 Worksho on IoT Enabling Technologies, and the ACM UbiComp'11 Workshop on Networking and Object Memories for IoT. He served as a Consultant with Asian Development Bank, Manila, Philippines, Bain \& Company, Boston, MA, USA, and KPMG, New York, NY, USA, and a peer reviewer for the Qatar National Research Foundation and the National Science Foundation in China. He is a member of the Association for Computing Machinery.

\end{IEEEbiography}

\begin{IEEEbiography}[{\includegraphics[width=1in,height=1.25in,clip,keepaspectratio]{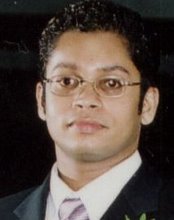}}]{Srimal Jayawardena}
received his BSc (Hons) in Electrical Engineering from the University of Peradeniya and his Bachelors in IT from the University of Colombo School of Computing, both with first class honours in 2004. He also obtained a Masters in Business Administration from the University of Moratuwa in 2009. He holds a PhD in Computer Science from The Australian National University, Canberra. He is currently working as post-doctoral research fellow at Computer Vision Laboratory (CI2CV),in CSIRO. His research interests include augmented reality, object recognition for the Internet of Things, computer vision, human computer interaction, and machine learning. He is a member of the Institute of Electrical and Electronics Engineers (IEEE).
\end{IEEEbiography}

\begin{IEEEbiography}[{\includegraphics[width=1in,height=1.25in,clip,keepaspectratio]{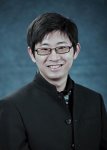}}]{Min Chen}
is currently a Full Professor with the School of Computer Science and Engineering, Huazhong University of Science and Technology, Wuhan, China. He was an Assistant Professor with Seoul National University, Seoul, Korea. He has authored over 150 papers. He serves as an Editor or Associate Editor of Wireless Communications and Mobile Computing, IET Communications, IET Networks, the International Journal of Security and Communication Networks (Wiley), the Journal of Internet Technology, the KSII Transactions on Internet and Information Systems, and the International Journal of Sensor Networks. He is the Managing Editor of the International Journal of Autonomous and Adaptive Communication Systems. He was the Co-Chair of the IEEE ICC 2012 Communications Theory Symposium and the IEEE ICC 2013 Wireless Networks Symposium. He was the General Co-Chair of the 12th IEEE International Conference on Computer and Information Technology in 2012. His research focuses on multimedia and communications, such as multimedia transmission over wireless network, wireless sensor networks, body sensor networks, RFID, ubiquitous computing, intelligent mobile agent, pervasive computing and networks, E-healthcare, medical application, machine to machine communications, and Internet of Things.
\end{IEEEbiography}

%%%%%%%% Template %%%%%%%%%%

%\begin{IEEEbiography}
%\end{IEEEbiography}

%\begin{IEEEbiographynophoto}

%%%%%%%%%%%%%%%%%%%%%%


%\vfill


\end{document}